\documentclass[aps,prd, twocolumn,nofootinbib]{revtex4}
\usepackage{graphicx}
\usepackage{amsmath, amsthm, amssymb}
\usepackage{slashed}
\usepackage{url}
\usepackage[pdftex]{hyperref}
\usepackage{color}
\usepackage{bm}

\newcommand{\qp}{\frac{q}{p}}
\newcommand{\qpabs}{\left|\frac{q}{p}\right|}
\newcommand{\pq}{\frac{p}{q}}
\newcommand{\pqabs}{\left|\frac{p}{q}\right|}
\newcommand{\sfermion}[1]{\tilde{\bar{#1}}}

\begin{document}
\vspace*{-1.7cm}

\begin{flushright}{\small FERMILAB-PUB-16-092-T}\end{flushright}

\vspace*{0.2cm}
\title{Baryogenesis via Particle--Antiparticle Oscillations}
\author{Seyda Ipek}
\affiliation{Fermi National Laboratory, Batavia, IL 60510, USA}
\author{John March-Russell}
\affiliation{Department of Physics, University of Oxford, Oxford, England}

\date{\today}

\begin{abstract}
$CP$ violation, which is crucial for producing the baryon asymmetry of the Universe, is enhanced in particle--antiparticle oscillations. We study particle--antiparticle oscillations (of a particle with mass $O(100~{\rm GeV})$) with $CP$ violation in the early Universe in the presence of interactions with $O$(ab--fb) cross-sections. We show that, if baryon-number-violating interactions exist, a baryon asymmetry can be produced via out-of-equilibrium decays of oscillating particles. As a concrete example we study a $U(1)_R$-symmetric, R-parity-violating SUSY model with pseudo-Dirac gauginos, which undergo particle--antiparticle oscillations. Taking bino to be the lightest $U(1)_R$-symmetric particle, and assuming it decays via baryon-number-violating interactions, we show that bino--antibino oscillations can produce the baryon asymmetry of the Universe.
\end{abstract}
\maketitle

\section{Introduction} \label{sec:intro}
There are more baryons than antibaryons in the Universe. Big Bang nucleosynthesis~\cite{Kneller:2004jz} and  cosmic microwave background~\cite{Ade:2015xua} measurements give the baryon asymmetry of the Universe
\begin{align}
\eta \equiv \frac{n_B-n_{\bar{B}}}{s}\simeq 10^{-10},
\end{align}
where $n_{B(\bar{B})}$ is the (anti)baryon number density and $s$ is the entropy density.

In order to explain this asymmetry three conditions must be met~\cite{Sakharov:1967dj}: \textbf{(i)} baryon number cannot be a conserved quantity, \textbf{(ii)} $C$ and $CP$ symmetries must be violated and \textbf{(iii)} baryon-number- and $CP$-violating processes should happen out of thermal equilibrium. Even though baryon number is anomalously violated at high temperatures, there is neither enough $CP$ violation nor an out-of-equilibrium process within the Standard Model (SM) to yield the observed baryon asymmetry. 

The baryon asymmetry of the Universe (BAU) is one of the strongest motivations for the need for physics beyond the SM. To some extent new physics models that deal with BAU can be divided into three types. \textbf{(1)} Extending the SM to include extra scalar particles can change the electroweak transition to a first-order phase transition, which provides out-of-equilibrium conditions. There can also be extra $CP$ violation in this extended Higgs sector. Two-Higgs-doublet models and many variants of supersymmetric models are the most studied examples of this type. \textbf{(2)} Extending the SM with heavy particles that decay out of equilibrium to SM particles. Examples include leptogenesis models with a  heavy right-handed neutrino. \textbf{(3)} A particle asymmetry can first be produced in a dark sector and then transferred to the SM sector. In these types of models, the origin of the asymmetry is often not studied due to a lack of understanding of the dark sector.\footnote{There are also models that produce the baryon asymmetry and a dark matter asymmetry through a common process.} For reviews on different types of genesis models, see, for example, \cite{Morrissey:2012db, Davidson:2008bu, Zurek:2013wia}.

In any baryogenesis scenario the origin of $CP$ violation is a crucial ingredient. $CP$ violation in scalar-extensions of the SM often generates large electric dipole moments (EDMs) for the elementary particles  which is highly constrained by null measurements of the electron EDM. (See, for example, \cite{Inoue:2014nva} for current EDM constraints in two-Higgs-doublet models.) In leptogenesis models $CP$ violation is attained by interference between tree-level and loop-level decays. 

A recently revived way of producing large $CP$ violation in order to explain the BAU is through particle--antiparticle oscillations. $CP$ violation can be enhanced in oscillations\footnote{$CP$ violation in oscillations exists only if there are both oscillations and decays~\cite{McKeen:2015cuz}.} if the decay width and the mass difference of the oscillating particles are comparable. If, in addition, these particles decay out of thermal equilibrium via baryon-/lepton-number-violating interactions, these decays can explain the observed baryon asymmetry. First studies  of particle oscillations as a source of baryon asymmetry (soft and resonant leptogenesis \cite{Pilaftsis:2003gt, Grossman:2003jv, D'Ambrosio:2003wy}) neglected the time evolution of $CP$ violation in the early Universe. Later it was shown that quantum effects~\cite{DeSimone:2007gkc} can be important for these scenarios~\cite{Fong:2008yv, Cirigliano:2007hb}. Detailed studies of flavor oscillations in soft/resonant leptogenesis models also showed that the time evolution of $CP$ violation is important to find the correct particle asymmetry~\cite{Cirigliano:2011di, Beneke:2010dz, Abada:2006fw}. However, these works still only included effects of the expansion of the Universe on particle oscillations: As long as the Hubble rate, $H(T)$, is larger than the oscillation frequency, $\omega_{\rm osc}$, particles do not have sufficient time to oscillate. Since the particle oscillations are suppressed, $CP$ violation, and hence the particle asymmetry, is also suppressed until $\omega_{\rm osc}>H(T)$. Another quantum process that suppresses oscillations is the \emph{quantum Zeno effect}~\cite{Misra:1976by}, also known as ``a watched pot never boils": Flavor-sensitive scatterings hinder oscillations. This effect was pointed out regarding neutrino oscillations in the early Universe~\cite{Dolgov:1980cq, Sigl:1992fn}, but was largely left out of particle--antiparticle oscillation discussions.\footnote{For studies of quantum decoherence effects in flavor oscillations in resonant leptogenesis, see, \emph{e.g.}, \cite{Dev:2014laa, Dev:2014wsa, Garbrecht:2014iia, Canetti:2012kh}} Refs.~\cite{Cirelli:2011ac, Tulin:2012re} incorporated elastic scatterings and annihilations in the analysis of  asymmetric dark matter oscillations. The effects of flavor-sensitive and flavor-blind interactions on particle--antiparticle oscillations were clearly identified in Ref.~\cite{Tulin:2012re} and cast out in the form of density matrix equations. (We point out that $CP$ violation was not considered in Ref.~\cite{Tulin:2012re}; since dark matter does not decay, there cannot be $CP$ violation in this system.)\footnote{As a way to evade all quantum decoherence effects, the authors of Ref.~\cite{Ghalsasi:2015mxa} used heavy particles that decay out-of-equilibrium at very low temperatures to mesinos. In that case there are no other processes that compete with oscillations and mesino--antimesino oscillations enhance $CP$ violation.} 

In this work we will study $CP$ violation in particle--antiparticle oscillations in the early Universe by studying the time evolution of the density matrix as outlined in Ref.~\cite{Tulin:2012re}.  Without any interactions, oscillations start when the expansion rate of the Universe drops below the oscillation rate of the particles, $H(T) < \omega_{\rm osc}$. If the particles interact with the relativistic plasma in the early Universe, the oscillations are further delayed until $\Gamma_{\rm int}<\omega_{\rm osc}$, where $\Gamma_{\rm int}$ is the rate of the interaction (and depends on the nature of the process). In order to enhance $CP$ violation in these oscillations, particles should oscillate at least a few times before they decay. The longer the oscillations are delayed the less $CP$ violation there is. (Since the start of oscillations is directly related to the baryon asymmetry in this scenario, we will address it extensively throughout the text.) We will show that a particle asymmetry can be produced via the oscillations and out-of-equilibrium decays of a particle of mass $O(100~{\rm GeV})$ with a mass splitting and decay rate of $O(10^{-6}~{\rm eV})$ even in the presence of interactions with $O$(ab--fb) cross-sections. As a specific example of this scenario, we will study a $U(1)_R$-symmetric supersymmetry (SUSY) model with R-parity violation. We will show that bino--antibino oscillations in this model can explain the measured baryon asymmetry. 

The rest of the paper is organized as follows. We start with a short review of particle--antiparticle oscillations for a pseudo-Dirac fermion in Section~\ref{sec:osc}. In Section~\ref{sec:oscearly} we study the oscillations of an electroweak scale pseudo-Dirac fermion in the early Universe (at temperature $T\sim O(10-100~{\rm GeV})$). We include interactions, specifically elastic scatterings with light particles and annihilations. In Section~\ref{sec:BAU} we calculate the baryon asymmetry that can be generated via the particle--antiparticle oscillations. We consider a specific example of this scenario in Section~\ref{sec:BAUreal}. We give our concluding remarks in Section~\ref{sec:summary}.

\section{Particle--Antiparticle Oscillations} \label{sec:osc}
In this section we briefly review particle--antiparticle oscillations. (For details, see~\cite{Ipek:2014moa}.) For simplicity let us focus on a single generation of pseudo-Dirac fermions with the mass Lagrangian
\begin{align}
-\mathcal{L}_{\rm mass}=M \chi\,\eta+\frac12 m_\chi\, \chi\,\chi+\frac12 m_\eta\, \eta\, \eta+{\rm h.c.}
\end{align}
where $\chi,\eta$ are two-component, left-handed Weyl fields, charged $+1,-1$ under a global $U(1)$, respectively. Let us define the Dirac field $\psi$,
\begin{align}
\psi=\left(\begin{array}{c} \eta_\alpha \\
					\chi^{\dagger\dot{\alpha}}\end{array}\right).
\end{align}
Particle and antiparticle states can be written in terms of the creation and annihilation operators
\begin{align*}
\psi(x)&=\sum_{s=\pm}\int\widetilde{dp} \left[ b_s(\mathbf{p})u_s(\mathbf{p})\,e^{ipx}+d_s^\dagger(\mathbf{p})v_s(\mathbf{p})\,e^{-ipx} \right], \\
\psi^c(x)& =\sum_{s=\pm}\int\widetilde{dp} \left[ d_s(\mathbf{p})u_s(\mathbf{p})\,e^{ipx}+b_s^\dagger(\mathbf{p})v_s(\mathbf{p})\,e^{-ipx} \right],
\end{align*}
where $\widetilde{dp}=\frac{d^3\mathbf{p}}{(2\pi)^3 2E_\mathbf{p}}$, such that
\begin{align*}
|\mathbf{p},s,\psi\rangle=d_s^\dagger(\mathbf{p})|0\rangle, \quad
|\mathbf{p},s,\psi^c\rangle=b_s^\dagger(\mathbf{p})|0\rangle.
\end{align*}
Given the Majorana masses $m_{\chi,\eta}$, particle and antiparticle states mix, and $\psi$ is called a \emph{pseudo-Dirac fermion}. In order to produce a baryon asymmetry, let us also consider the following effective operators that violate baryon or lepton number
\begin{align}
-\mathcal{L}_{\rm int}=g_{\chi}\,\chi B X+g_\eta\, \eta B X + g_\chi'\,\chi L Y+g_\eta'\,\eta L Y+{\rm h.c.}, \label{eq:Lint}
\end{align}
where $B/L$ are states with +1 baryon/lepton number. $X,Y$ are states with zero baryon and lepton number and are given by the details of the model. The effective coupling constants $g,g'$ have the proper dimensions to make the Lagrangian dimension four. If they are heavy enough, $\psi$-particles and antiparticles can decay via these interactions to baryons or leptons. 

Including the mass terms and focusing only on the baryon-number-violating interactions, the Hamiltonian is 
\begin{align}
\mathbf{H}&=\mathbf{M}-\frac{i}{2}\mathbf{\Gamma} \label{eq:hamiltonian},
\end{align}
with
\begin{align}
\mathbf{M}&=\left(\begin{array}{cc}
								M_D & M_M \\
								M_M^\ast & M_D
								\end{array} \right),\notag \\
\mathbf{\Gamma}&\simeq \gamma \left(\begin{array}{cc}
								|g_\chi|^2+|g_\eta|^2 & 2\,g_\chi^\ast g_\eta\\
								2\,g_\chi g_\eta^\ast & |g_\chi|^2+|g_\eta|^2
								\end{array} \right).
\end{align}
The masses $M_D$ and $M_M$ are the renormalized masses. The Dirac mass is multiplicatively renormalized from its tree-level value, $M_D\simeq M$. The loop contributions to the tree-level Majorana mass, $m=\frac{m_\chi+m_\eta^\ast}{2}$, are proportional to the Dirac mass $M$  and the $U(1)$-violating interaction coefficients. $\gamma$ is also proportional to the Dirac mass and is given by the details of the model. 

Note that even if the tree-level Majorana masses are zero, they will be generated at loop level via the interactions in Eq.~\ref{eq:Lint}. The loop contribution to the Majorana mass $\delta\sim C |g_\chi g_\eta^\ast| M$, where $C$, depending on the details of the model, can be a dimensionful coefficient which also incorporates the loop factors. On the other hand the decay width $\Gamma\sim C' (|g_\chi|^2+|g_\eta|^2)M$, where $C'$ is a coefficient that has the appropriate dimensions and phase-space factors. In this work, for simplicity, we assume a hierarchy of couplings $|g_\chi|\gg |g_\eta|$. Hence the loop contributions to the Majorana masses are smaller than the decay width of the particles. (For example in~\cite{Ipek:2014moa} it was shown that for a Yukawa-type interaction, $\delta/\Gamma\sim |g_\eta|/|g_\chi|$.) Then, if the tree-level Majorana masses were zero, one would expect the loop-induced Majorana mass $M_M < \Gamma$. Later we will show that in order to have large $CP$ violation, we need $M_M\sim \Gamma$. Hence, we take the tree-level Majorana mass, $m$, as a free parameter and require that it is greater than the loop contributions, such that $M_M \simeq m$.\footnote{Interactions with states that are heavier than $\psi$ could contribute to the Majorana masses while not changing the decay rate. Hence raising the loop-induced Majorana mass relative to the decay rate. There can also be models where $M_M\sim \Gamma$ \emph{naturally}. We leave this as a model building exercise.}

The eigenstates of the Hamiltonian in Eq.~\ref{eq:hamiltonian} are
\begin{align*}
|\psi_H\rangle =p|\psi\rangle-q|\psi^c\rangle, \quad
|\psi_L\rangle =p|\psi\rangle+q|\psi^c\rangle,
\end{align*}
with eigenvalues $\omega_{H,L}$, where $H,L$ refer to heavy and light states respectively. We also have
\begin{align*}
\left(\qp\right)^2=\frac{M_{12}^\ast-(i/2)\Gamma_{12}^\ast}{M_{12}-(i/2)\Gamma_{12}}.
\end{align*}
Mass and width differences between the heavy and light eigenstates are defined as
\begin{align*}
\Delta m&=m_H-m_L= \Re(\omega_H-\omega_L), \\
\Delta\Gamma&=\Gamma_H-\Gamma_L=-2\Im(\omega_H-\omega_L),
\end{align*}
where 
\begin{align*}
\omega_H-\omega_L=2\sqrt{\left(M_{12}^\ast-\frac{i}{2}\Gamma_{12}^\ast\right)\left(M_{12}-\frac{i}{2}\Gamma_{12}\right)}.
\end{align*}
The total width of the states is defined as $\Gamma=\frac{\Gamma_H+\Gamma_L}{2}$.

One can always rotate two linear combinations of $\chi$ and $\eta$ to make $M$ and $m$ real. We can also rotate  $B X$ to make $g_\chi$ or $g_\eta$ real, but not necessarily both at the same time. Hence it is possible to have a phase difference between $\mathbf{M}$ and $\mathbf{\Gamma}$, which will be a source of \emph{CP} violation. From now on we will assume the mass matrix is real, and put the relative phase in the decay matrix.

Assuming  $r=\frac{|g_\eta|}{|g_\chi|}\ll 1$, we can write
\begin{align}
\mathbf{\Gamma} \simeq \Gamma \left(\begin{array}{cc}
														1 & 2r e^{i\phi_\Gamma} \\
														2r e^{-i\phi_\Gamma} & 1
\end{array} \right),
\end{align}
where
\begin{align}
\Gamma=(|g_\chi|^2+|g_\eta|^2)\,\gamma.
\end{align}

In this approximation the oscillation parameters are given by
\begin{align}
& x\equiv \frac{\Delta m}{\Gamma}\simeq\frac{2m}{\Gamma}, \quad\quad\quad y\equiv \frac{\Delta\Gamma}{2\Gamma}\simeq 2r\cos\phi_\Gamma,\notag \\
&\qpabs\simeq 1-\frac{2r}{x}\sin\phi_\Gamma, \quad \beta \equiv \arg\left(\qp\frac{g_\eta}{g_\chi}\right) \simeq\phi_\Gamma \pm \pi.
\end{align}

The time evolution of a state that is purely a $|\psi\rangle$ or $|\psi^c\rangle$ at $t=0$ is
\begin{align}
|\psi(t)\rangle&=g_+(t)|\psi\rangle-\qp g_-(t)|\psi^c\rangle, \notag \\
|\psi^c(t)\rangle&=g_+(t)|\psi^c\rangle-\pq g_-(t)|\psi\rangle, \label{eq:timeevolv}
\end{align}
where
\begin{align}
g_\pm(t) = \frac12\left( e^{-im_Ht-\frac{1}{2}\Gamma_Ht}\pm e^{-im_Lt-\frac{1}{2}\Gamma_Lt}\right). \label{eq:gpm}
\end{align}

\subsection{$CP$ Violation}\label{sec:cpviolation}
$CP$ violation can be enhanced in particle--antiparticle oscillations. We quantify the $CP$ violation that is important for baryogenesis as a single particle asymmetry,
\begin{align}
\epsilon&=\int_0^\infty dt\, \frac{\Gamma(\psi/ \psi^c\to B)- \Gamma(\psi / \psi^c\to\bar{B})}{\Gamma(\psi/\psi^c\to B) + \Gamma(\psi/\psi^c\to\bar{B})},
\end{align}
where $B$ and $\bar{B}$ refer to baryon and antibaryon final states respectively. (Defining a lepton asymmetry through the lepton-number-violating terms in $\mathcal{L}_{\rm int}$ is straightforward.) $\Gamma(\psi/ \psi^c\to B)$ is the time-dependent decay rate for an initially pure-$|\psi\rangle$ or $|\psi^c\rangle$ state to decay to a baryon final state. Time integration is distributed over each decay rate.

\begin{figure}[h!]
\includegraphics[width=\linewidth]{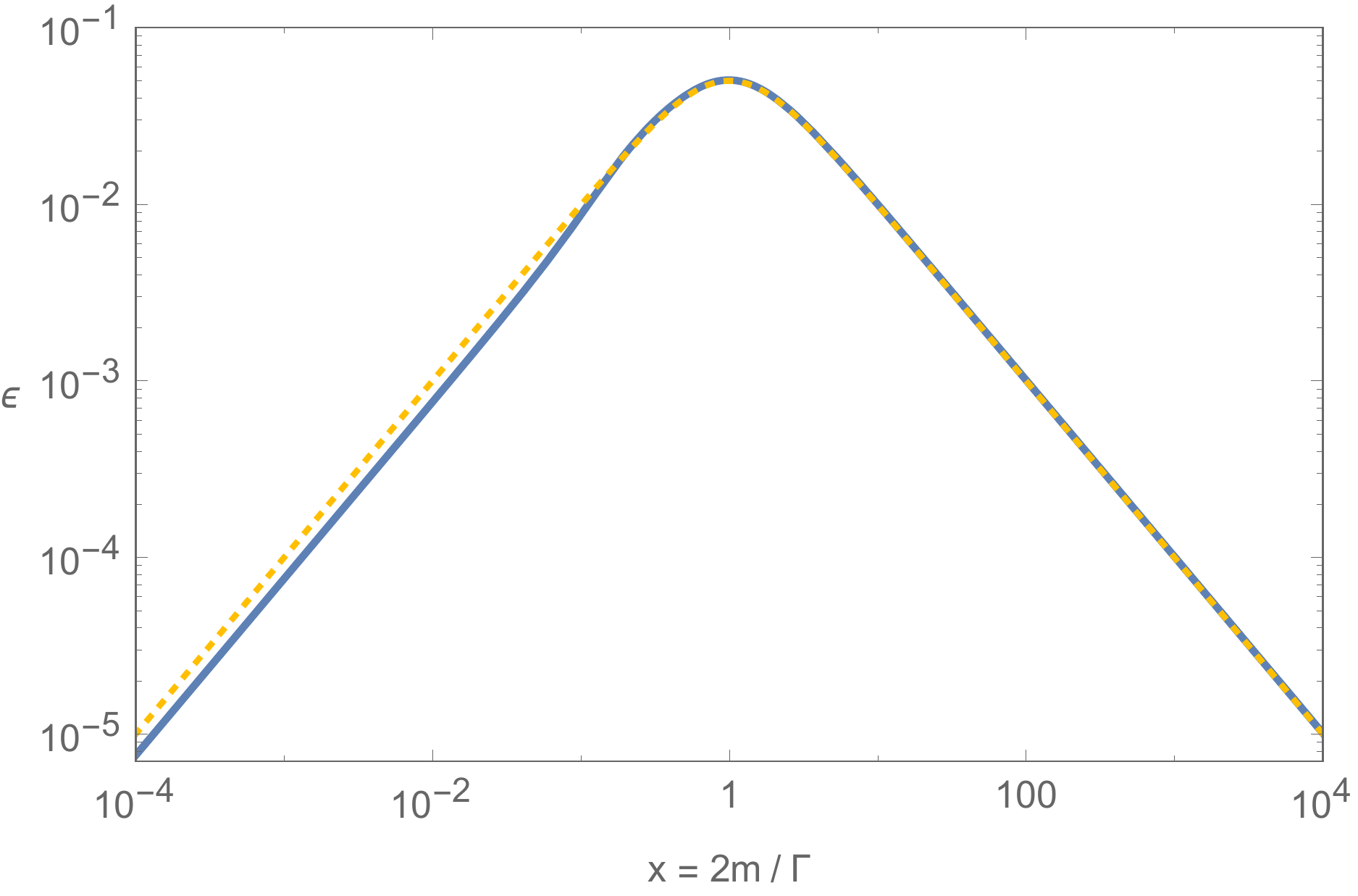}
\caption{Amount of $CP$ violation as defined in Eq.~\ref{eq:epsilon} (blue, thick) and the approximation in Eq.~\ref{eq:epsapprox} (orange, dashed). We use $r=0.1,~ \sin\phi_\Gamma=0.5$.} \label{fig:epsilon}
\end{figure}

Using the results of the previous section, $CP$ violation becomes
\newpage
\begin{widetext}
\begin{align}
\epsilon&= \frac{\int_0^\infty dt\,\left(\pqabs^2-\qpabs^2\right)|g_-|^2(1-r^2) }{\int_0^\infty dt\,\left(\left[2|g_+|^2+\left(\pqabs^2+\qpabs^2\right)|g_-|^2\right](1+r^2) -4r\,\Re\left[ g_+^\ast g_-\left(\qpabs e^{i\beta}+\pqabs e^{-i\beta}\right) \right]\right)} .\label{eq:epsilon}
\end{align}
\end{widetext}
See Appendix~\ref{sec:timedepdec} for details.  There is no $CP$ violation for $r=1$ or $|q/p|=1$.  For $r<1$ and $x\geq1$, the $CP$ violation can be approximated as
\begin{align}
\epsilon\simeq \frac{2\,x\,r \sin\phi_\Gamma }{1+x^2}. \label{eq:epsapprox}
\end{align}
As can be seen in Fig.~\ref{fig:epsilon}, $CP$ violation is maximized for $x\sim 1$, i.e. $\Delta m\sim \Gamma$. 

\section{Oscillations in the early universe} \label{sec:oscearly}

Particle--antiparticle oscillations in the early Universe are different than those expected to be seen at colliders. In an expanding Universe with a dense and relativistic plasma, the dynamics are defined by a competition between rates for many processes. For example, even in the absence of any interactions with the plasma,  oscillations do not start as long as $\omega_{\rm osc}\simeq 2m<H(T)$. Even after the Hubble rate drops below the oscillation rate, the particles and antiparticles might be interacting with the plasma such that the states decohere. If, for example, only the particle states were scattering with the plasma, elastic scatterings could keep this state from oscillating into an antiparticle state. However elastic scatterings that do not differentiate between particles and antiparticles do not cause decoherence as discussed in~\cite{Tulin:2012re}.

Other complications with this early Universe study are possible finite temperature effects. We will show later that the baryon number production happens at temperatures much less than the $\psi$-particle mass, $T\ll M$. Hence, we do not expect thermal corrections to the particle mass to be important. We also ignore thermal corrections (\emph{e.g.} due to finite-temperature SM fermion masses) to the decay rate. Possibly the most important finite-temperature effect is that on the Majorana mass, hence the mass difference between the $\psi$-particle mass eigenstates. Barring interactions beyond those in Eq.~\ref{eq:Lint}, these corrections will be proportional to the temperature and a combination of the coupling constants. Since the renormalization corrections to the mass difference are proportional to $M$, and since $T\ll M$ in the region of interest, we ignore thermal corrections to the mass difference.

In Refs.~\cite{Cirelli:2011ac, Tulin:2012re} effects of oscillations in an asymmetric DM scenario were studied. There are many similarities we can draw from that picture, and a few differences, namely that in our case, the particles/antiparticles decay allowing $CP$ violation in oscillations. In general, when oscillations, annihilations and scatterings are present, the relevant Boltzmann equations that define the particle number densities ($Y\equiv n/s\propto \sum_{\psi,\psi^c}f_{ij}|\psi_i\rangle\langle\psi_j |$, with the generalized quantum distribution functions $f_{ij}$) are written in a density matrix form as\footnote{This equation describes nonrelativistic particles (the Hamiltonian is given at $\mathbf{p}=0$).}
\begin{align}
zH\frac{d\mathbf{Y}}{dz}&=-i\bigr(\mathbf{H}\mathbf{Y}-\mathbf{Y}\mathbf{H}^\dagger\bigl)-\sum_{+,-}\frac{\Gamma_\pm}{2}[O_\pm,[O_\pm,\mathbf{Y}]]  \label{eq:oscBoltzmann} \\
&\qquad-\sum_{+,-}s\langle\sigma v\rangle_\pm\left(\frac12 \{\mathbf{Y},O_\pm \bar{\mathbf{Y}}O_\pm\}-Y_{\rm eq}^2\right),\notag 
\end{align}
where $z=M/T$, $s=\frac{2\pi^2}{45}g_\ast (T) T^3$ is the entropy density and $g_\ast(T)$ is the effective number of relativistic degrees of freedom at temperature $T$ (we take $g_\ast\sim100$ for temperatures $O$(100~GeV)).  $\Gamma_+(\Gamma_-)$ is the elastic scattering rate that is flavor blind (sensitive) where ``flavor" refers to particle/antiparticle nature (see below), $\langle\sigma v\rangle$ is the annihilation rate and $O_\pm={\rm diag}(1,\pm 1)$. (Note that both types of interactions can be present.) The density matrix and the Hamiltonian are
\begin{align}
&\mathbf{Y}=\left(\begin{array}{cc}
		Y_{\psi}&	Y_{\psi\psi^c}\\
		Y_{\psi^c\psi}&	Y_{\psi^c}
	\end{array}\right),\quad  \bar{\mathbf{Y}}=\left(\begin{array}{cc}
								Y_{\psi^c}&	Y_{\psi\psi^c}\\
								Y_{\psi^c\psi}&	Y_{\psi}
								\end{array}\right), \notag\\
&Y_{\rm eq}(z)\simeq
    \begin{cases}
      \frac{135\zeta(3)}{4\pi^4 g_\ast}, & z<1 \\
      \frac{45}{2\pi^3\sqrt{2\pi}g_\ast}z^{3/2}e^{-z}, & z>1,
    \end{cases} \notag \\   
&\mathbf{H}=\left(\begin{array}{cc}
		M-i\Gamma/2&	m-i\Gamma\, r\,e^{i\phi_\Gamma}\\
		m-i\Gamma \,r\,e^{-i\phi_\Gamma}&	M-i\Gamma/2
		\end{array}\right).
\end{align}
The first term in Eq.~\ref{eq:oscBoltzmann} describes oscillations (and decays), the second term elastic scatterings and the third term  annihilations.  

The nature of elastic scatterings and annihilations depends on the details of the model. We will study a specific model in Section~\ref{sec:BAUreal}. For now  let us consider a generic effective 4-fermion operator,
\begin{align}
-\mathcal{L}_{\rm scat} = \frac{1}{\Lambda^2}\bar{\psi}\, \Gamma^a\, \psi\, \bar{f}\, \Gamma_b\, f, \label{eq:Lscat}
\end{align}
where $f$ is a light fermion, $\Lambda$ is the interaction scale, and $\Gamma^a=\{1,\gamma^\mu,\gamma_5,\gamma^\mu\gamma_5,\sigma^{\mu\nu}= \frac12[\gamma^\mu,\gamma^\nu]\}$ gives the gamma-matrix structure of the interaction. This effective operator gives rise to both elastic scatterings ($\psi f\to\psi f$) and annihilations ($\psi \psi^c\to f\bar{f}$). Under the transformation $\psi\to\psi^c$, flavor-blind and flavor-sensitive interactions are defined as
\begin{align*}
\mathcal{L}_{\rm scat}\to
	\begin{cases}
	+ \mathcal{L}_{\rm scat},\quad & \text{flavor blind} \\
	- \mathcal{L}_{\rm scat},\quad & \text{flavor sensitive}. 
	\end{cases}
\end{align*}
If the interactions are flavor blind, $O_+$ is the identity matrix and the second term in Eq.~\ref{eq:oscBoltzmann} is identically zero. Hence flavor-blind scatterings do not cause decoherence. However, as we will see, oscillations are delayed due to flavor-blind annihilations.

\begin{figure}[h!]
\includegraphics[width=\linewidth]{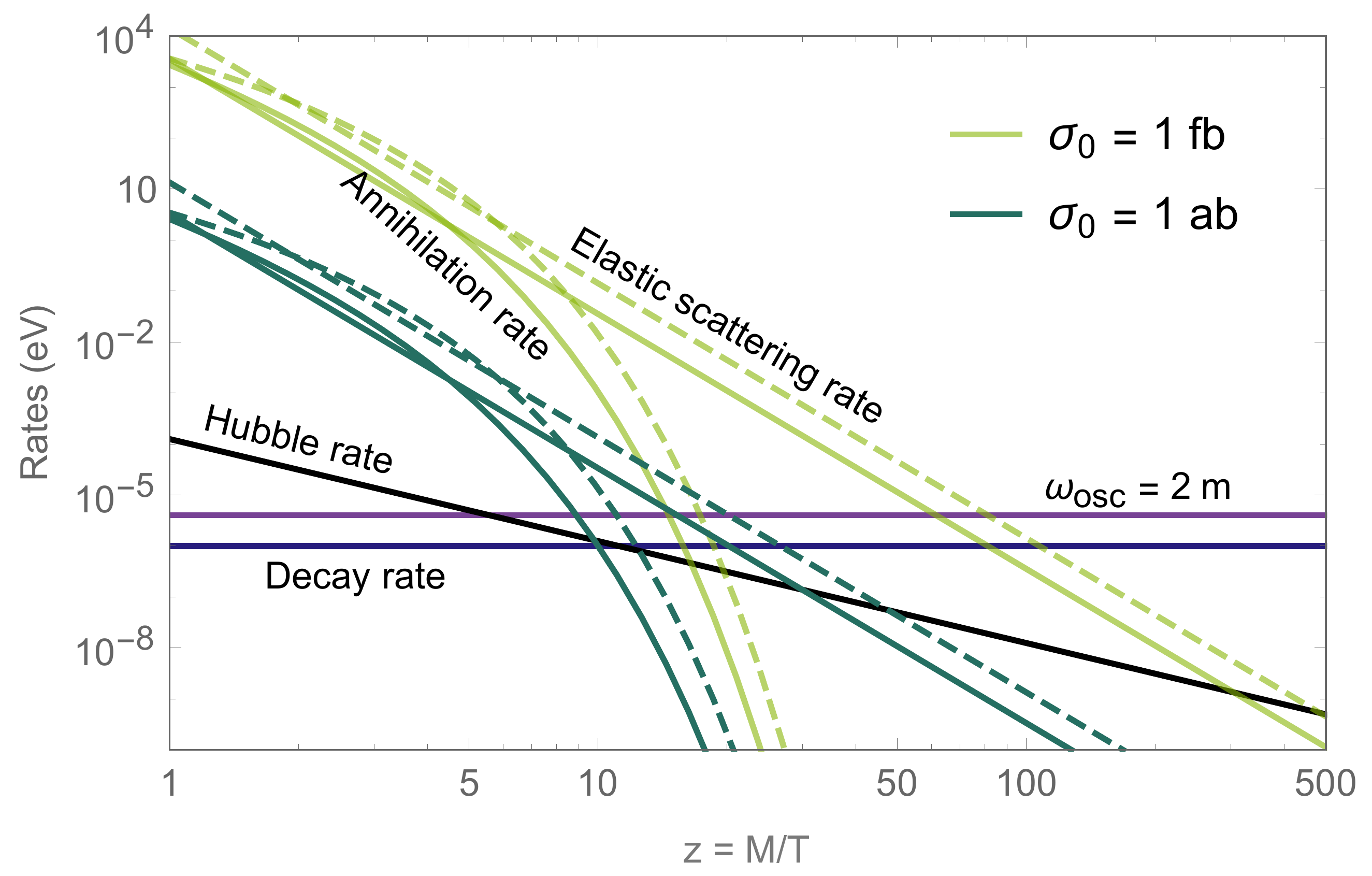}
\caption{Comparison of the decay rate, the Hubble rate, the oscillation frequency and the annihilation and the elastic scattering rates for scalar (solid) and vector (dashed) interactions for $M=300$~GeV, $m=2\times10^{-6}~$eV, $\Gamma= 10^{-6}$~eV, $\sigma_0=1~$fb (light), and $\sigma_0=1~$~ab (dark).}\label{fig:ratesvz}
\end{figure}

In Fig.~\ref{fig:ratesvz} we compare some representative rates for several processes relevant for net particle number production in the early Universe. These are given as follows.
\begin{enumerate}
\item The expansion rate of the Universe is
\begin{align}
H(T)=\sqrt{\frac{4\pi^3g_\ast}{45}}\frac{T^2}{M_{\rm pl}},
\end{align}
where  $M_{\rm pl}\simeq 1.2\times 10^{19}~$GeV is the Planck mass. 

\item The decay rate $\Gamma$ is determined by requiring that the particles decay out of equilibrium:
\begin{align}
\Gamma < H(T\sim M) \simeq 10^{-4}~{\rm eV},
\end{align}
for $M=300~$GeV. We take $\Gamma=10^{-6}$~eV (such that the $\psi$-particles decay at $T\simeq 30$~GeV) as a benchmark value.

\item The oscillation rate $\omega_{\rm osc}$ is set by the mass splitting between the heavy and light mass eigenstates. For pseudo-Dirac fermions $\omega_{\rm osc}\sim2m$ where $m=\frac{m_\chi+m_\eta}{2}$ is the Majorana mass of the fermions. ($CP$ violation in oscillations is maximized for $2m\sim \Gamma$.) Oscillations cannot proceed before the Hubble rate drops below $\omega_{\rm osc}$. Neglecting scatterings and annihilations, for particles of mass 300~GeV with a mass splitting of $2\times10^{-6}~{\rm eV}$  the onset of oscillations is delayed until $z\sim 6$  (see Fig.~\ref{fig:ratesvz}).

\item Elastic scatterings and annihilations affect oscillations. For a general study, we consider the Lagrangian in Eq.~\ref{eq:Lscat} with \textbf{(i)} scalar interactions, $\Gamma^{a,b}=1$, which are flavor blind and \textbf{(ii)} vector interactions, $\Gamma^{a,b}=\gamma^\mu$, which are flavor sensitive.\footnote{In Section~\ref{sec:BAUreal} we will consider binos scattering with SM fermions via a sfermion exchange, which corresponds to $\Lambda\propto m_{\rm sf}^2$ with both flavor-blind and flavor-sensitive interactions.} For non-relativistic $\psi$-particles and assuming $m_f=0$, these interactions give the following thermally averaged annihilation and scattering cross sections.
\begin{align}
\mathbf{(i)}~\langle\sigma_{\rm ann}^S v\rangle&=\frac{3N M T }{4\pi\Lambda^4},\quad \langle\sigma_{\rm scat}^Sv\rangle=\frac{N T^2}{4\pi\Lambda^4},\notag \\
\mathbf{(ii)}~\langle\sigma_{\rm ann}^V v\rangle&=\frac{N M^2 }{\pi\Lambda^4},\qquad \langle\sigma_{\rm scat}^V v\rangle=\frac{N T^2}{\pi\Lambda^4}, \label{eq:x-sec}
\end{align}
where $v$ is the relative velocity of incoming particles and $N$ accounts for degrees of freedom, \emph{e.g.} color factors. Annihilations and scatterings can be enhanced significantly if there are many fermions with interactions as in Eq.~\ref{eq:Lscat}.

In Fig.~\ref{fig:ratesvz} we plot the following annihilation and scattering rates per particle (for $M=300~$GeV).
\begin{align}
\mathbf{(i)}~\quad \Gamma_{\rm ann}^S&=n_\psi \langle\sigma_{\rm ann}^S v\rangle=\frac{3M^3}{4\pi\sqrt{2\pi}}\frac{e^{-z}}{z^{5/2}} \sigma_0,\notag \\
\Gamma_{\rm scat}^S&=n_f \langle\sigma_{\rm scat}^S v\rangle = \frac{3\zeta(3)M^3}{8\pi^2 z^5}\sigma_0,\notag \\
\mathbf{(ii)}~\quad \Gamma_{\rm ann}^V&=n_\psi \langle\sigma_{\rm ann}^V v\rangle=\frac{M^3}{\pi\sqrt{2\pi}}\frac{e^{-z}}{z^{3/2}} \sigma_0,\notag \\
\Gamma_{\rm scat}^V&=n_f \langle\sigma_{\rm scat}^V v\rangle = \frac{3\zeta(3)M^3}{2\pi^2z^5}\sigma_0, \label{eq:rates}
\end{align}
where $\sigma_0=\frac{N M^2}{\pi\Lambda^4}$ is an effective cross section and $n_{\psi,f}$ are (equilibrium) number densities. The annihilation rate through a scalar operator is velocity suppressed compared to the vector case. Annihilations are Boltzmann suppressed compared to scatterings off light fermions. 
\end{enumerate}

Unless otherwise noted, we use the following parameters throughout this paper:
\begin{align}
M&=300~{\rm GeV},~ m=2\times 10^{-6}~{\rm eV},~\Gamma=10^{-6}~{\rm eV},\notag\\
~r&=0.1,\qquad\quad\sin\phi_\Gamma=0.5, \qquad~~\sigma_0={\rm ab - fb}. \label{eq:parameters}
\end{align}
These parameters are chosen as benchmark values. They are by no means \emph{fine-tuned}. The particle asymmetry, proportional to the $CP$ violation parametrized by $\epsilon$, can be made larger by an $O(1)$ amount by changing $r$ and $\sin\phi_\Gamma$. Mass of the $\psi$-particles can be $O$(TeV) or higher. Even though one needs $m\sim \Gamma$ to maximize the $CP$ violation, an asymmetry as small as $10^{-10}$ would not need $O(1)$ $CP$ violation. We will show later that mass differences as large as $O$(eV) can produce enough baryon asymmetry. The out-of-equilibrium condition puts an upper bound on the decay width, $\Gamma\lesssim 10^{-4}~$eV for $M=300~$GeV. We use $\Gamma=10^{-6}~$eV due to an interesting collider signature: if produced, particles with a width of $10^{-6}~$eV travel $\sim\! 20~$cm at a collider before they decay, giving rise to displaced-vertex signatures. (A detailed study of using  displaced vertices to probe baryogenesis was given in Ref.~\cite{Cui:2014twa}.)  After setting this width, a mass difference of $2\times 10^{-6}~$eV is chosen to make the oscillations more visible.  For effective cross sections larger than a fb, it is very hard to produce enough asymmetry. 

\subsection{Toy Case: Oscillations and Decays} \label{sec:toycase}
Let us first ignore annihilations and elastic scatterings in Eq.~\ref{eq:oscBoltzmann} and study oscillations and decays in an expanding Universe.  The Boltzmann equations in this case read
\begin{align}
\frac{d\Delta(z)}{dz}&=-\frac{\Gamma}{zH}\,\Delta(z)+i\frac{2m}{zH}\,\Xi(z) ,\notag \\
\frac{d\Sigma(z)}{dz}&=-\frac{\Gamma}{zH}\,\Sigma(z)-\frac{2\Gamma\,r}{zH}\bigl[\cos\phi_\Gamma\Pi(z)-i\sin\phi_\Gamma\Xi(z)\bigr], \notag \\
\frac{d\Xi(z)}{dz}&=-\frac{\Gamma}{zH}\,\Xi(z)+i\frac{2m}{zH}\,\Delta(z)-i\frac{2\Gamma\,r\sin\phi_\Gamma}{zH} \Sigma(z),\notag\\
\frac{d\Pi(z)}{dz}&=-\frac{\Gamma}{zH}\,\Pi(z)-\frac{2\Gamma\,r\cos\phi_\Gamma}{zH}\Sigma(z),\label{eq:Boltzoscdecre}
\end{align}
where we defined
\begin{align}
\Delta (z)&=Y_\psi-Y_{\psi^c},\quad\quad \Xi(z)=Y_{\psi\psi^c}-Y_{\psi^c\psi}, \notag \\
\Sigma(z)&=Y_\psi+Y_{\psi^c},\quad\quad\Pi(z)=Y_{\psi\psi^c}+Y_{\psi^c\psi}. \label{eq:partsums}
\end{align}

We can solve for $\Xi(z)$ from the first  equation and plug it into the third equation. Then, with a change of variables $y=z^2$, we have
\begin{align}
\frac{d^2\Delta(y)}{dy^2}+2\,\xi\,\omega_0\,\frac{d\Delta(y)}{dy}+\omega_0^2\,\Delta(y)=-\epsilon\,\omega_0^2\,\Sigma(y), \label{eq:Delta}
\end{align}
where 
\begin{align*}
\omega_0^2=\frac{\Gamma^2+4m^2}{4[z^2H(z)]^2},\quad \xi=\frac{\Gamma}{\sqrt{\Gamma^2+4m^2}}.
\end{align*}
$\epsilon$ is the measure of $CP$ violation given in Eq.~\ref{eq:epsapprox}. This differential equation can be solved analytically for zero and nonzero $CP$ violation. 
\begin{figure*}
\includegraphics[width=\textwidth]{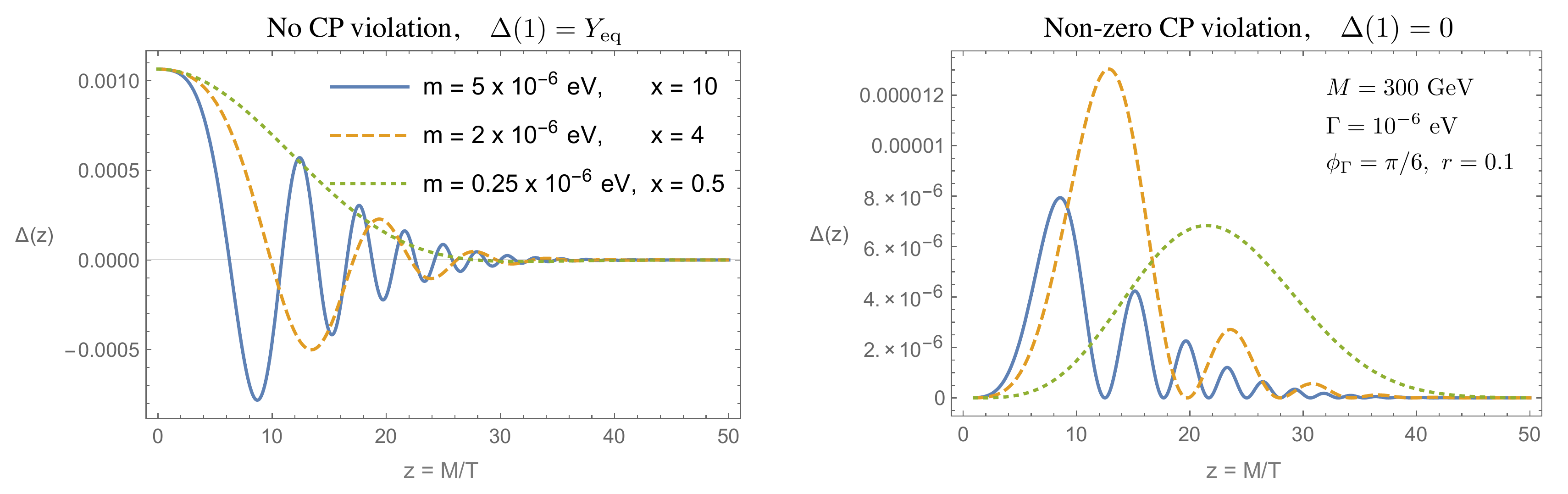}
\caption{The $\psi$-asymmetry for different values of the mass splitting and for $M=300$~GeV, $\Gamma=10^{-6}$~eV without annihilations or scatterings. For smaller values of $\omega_{\rm osc}=2m$, oscillations are delayed. \textbf{(Left)} Without any $CP$ violation, and $\Delta(1)=Y_{\rm eq}$. Since there is no $CP$ violation, no asymmetry is produced if the initial conditions are symmetric. \textbf{(Right)} Symmetric initial conditions, $\Delta(1)=0$, with nonzero $CP$ violation, $r=0.1,\sin\phi_\Gamma=0.5$. }\label{fig:oscdec}
 \end{figure*}

\textbf{(1) Without $CP$ violation.} For $\epsilon=0$, the above differential equation describes a damped oscillator with the solution
\begin{align}
\Delta(z)=A \exp\left(-\frac{\Gamma}{2H(z)}\right)\cos\left(\frac{m}{H(z)}+\delta\right),
\end{align}
where $A$ and $\delta$ are determined by initial conditions. Note that if there is no $CP$ violation and if the initial conditions are symmetric ($\Delta(0)=0$), $\Delta(z)$ stays zero for all $z$ and there is no net $\psi$-number production as expected. In Fig.~\ref{fig:oscdec} we plot $\Delta(z)$ for asymmetric initial conditions, $\Delta(1)=Y_{\rm eq}$. One can also see in Fig.~\ref{fig:oscdec} that for smaller Majorana masses (hence smaller oscillation frequency, since $\omega_{\rm osc}\simeq 2m$) the onset of oscillations is delayed. (We use oscillations in $\Delta(z)$ as a proxy for particle--antiparticle oscillations.) The time when  oscillations start, $z_{\rm osc}$, can be approximated from $2m\sim H(z_{\rm osc})$ as 
\begin{align}
z_{\rm osc}\sim 6\, \sqrt{\frac{2\times 10^{-6}~{\rm eV}}{m}}\left(\frac{M}{300~{\rm GeV}}\right). \label{eq:zoscH}
\end{align}

\textbf{(2) With $CP$ Violation.} For nonzero $\phi_\Gamma$ and $r$, Eq.~\ref{eq:Delta} describes a damped-driven oscillator, where $\Sigma(z)$ plays the role of a driving force. To get an analytic solution in this case we assume that particles and antiparticles are produced with equilibrium number densities, hence $\Sigma(1)\simeq2Y_{\rm eq}(1)$. In the absence of annihilations or scatterings, the total number density decays exponentially with the decay rate $\Gamma$. Thus we take
\begin{align}
\Sigma(z)=2\,Y_{\rm eq}(1) \exp\left( -\frac{\Gamma}{2H(z)} \right)\quad \text{for}\quad z>1.
\end{align}
With this driving force, Eq.~\ref{eq:Delta} can be solved analytically
\begin{align}
\Delta (z)\simeq A \,\epsilon Y_{\rm eq}(1) \exp\left( -\frac{\Gamma}{2H(z)}\right)\sin^2\left(\frac{m}{2H(z)}+\delta\right),
\end{align}
where again  $A$ and $\delta$ are determined by initial conditions. Note that in this case, where there is $CP$ violation, a nonzero asymmetry is produced even with symmetric initial conditions and it is proportional to the $CP$ violation parameter $\epsilon$. This $\psi$-asymmetry is plotted for different Majorana masses in Fig.~\ref{fig:oscdec}. For $m\gtrsim \Gamma$ the $\psi$-particles oscillate  a few times before decaying and the asymmetry is enhanced. In order to get a large asymmetry as well as making the oscillations more apparent, we use $m=2\times10^{-6}~$eV in the rest of our analysis.

\subsection{Flavor-Blind Interactions} \label{sec:flavor-blind}
Let us now include scalar interactions, which are flavor blind. As discussed earlier, flavor-blind elastic scatterings do not affect the oscillations. However flavor-blind annihilations change the  Boltzmann equations in Eq.~\ref{eq:Boltzoscdecre}:
\begin{widetext}
\begin{align}
\frac{d\Delta(z)}{dz}&=-\frac{\Gamma}{zH(z)}\,\Delta(z)+i\frac{2m}{zH(z)}\,\Xi(z) ,\notag \\
\frac{d\Sigma(z)}{dz}&=-\frac{\Gamma}{zH(z)}\,\Sigma(z)-\frac{2\Gamma\,r}{zH(z)}\bigl[\cos\phi_\Gamma\Pi(z)-i\sin\phi_\Gamma\Xi(z)\bigr]-\frac{s(z)\langle \sigma_{\rm ann}^S v\rangle(z)}{2\,zH(z)}\biggl(\Sigma^2(z)-\Delta^2(z)+\Pi^2(z)-\Xi^2(z)-4Y_{\rm eq}^2(z)\biggr),\notag \\
\frac{d\Xi(z)}{dz}&=-\frac{\Gamma+s(z)\langle \sigma_{\rm ann}^S v\rangle(z)\,\Sigma(z)}{zH(z)}\,\Xi(z)+i\frac{2m}{zH(z)}\,\Delta(z)-i\frac{2\Gamma\,r\sin\phi_\Gamma}{zH(z)} \Sigma(z), \notag\\
\frac{d\Pi(z)}{dz}&=-\frac{\Gamma+s(z)\langle \sigma_{\rm ann}^S v\rangle(z)\,\Sigma(z)}{zH(z)}\,\Pi(z)-\frac{2\Gamma\,r\cos\phi_\Gamma}{zH(z)}\Sigma(z), \label{eq:Bflavorblind}
\end{align}
\end{widetext}
where the thermally averaged annihilation cross section $\langle \sigma_{\rm ann}^S v\rangle(z)$ is given in Eq.~\ref{eq:x-sec}. We made the $z$ dependence of each term explicit in the above equations. For scalar interactions we have $\langle\sigma v\rangle=\sigma_0/z$ with $\sigma_0$ constant. (Numerical solutions to these equations are given in Fig.~\ref{fig:sigmadelta}.)

\begin{figure*}[!htb]
\includegraphics[width=\textwidth]{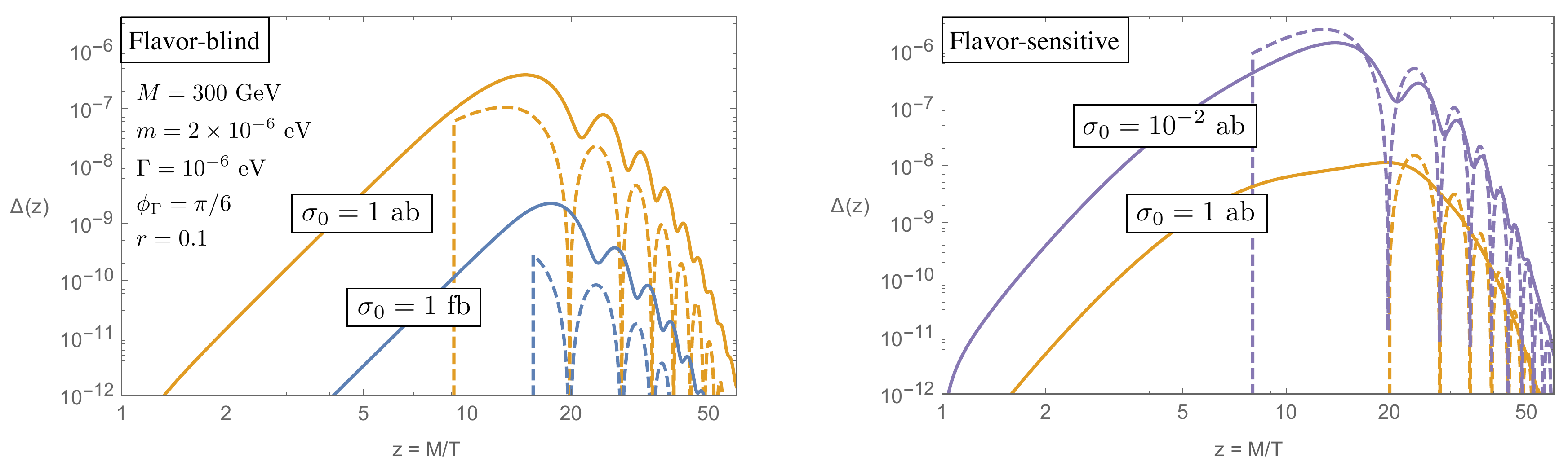}
\caption{Comparison of the numerical solutions to the Boltzmann equations (solid) and the approximations given in the text (dashed) for the $\psi$-asymmetry $\Delta(z)$ in the presence of \textbf{(left)} flavor-blind and \textbf{(right)} flavor-sensitive interactions. For the numerical solutions we use the initial condition $\Delta(1)=0$, while for the analytical approximation we use $\Delta(z<z_{\rm osc})=0$. The following parameters are used in both plots: $M=300$~GeV, $\Gamma=10^{-6}$~eV, $m=2\times 10^{-6}$~eV, $r=0.1$, $\sin\phi_\Gamma=0.5$.} \label{fig:deltaapp}
 \end{figure*}

The density equations no longer have closed-form solutions. However we can still make some comments without a numerical solution and understand the oscillation behavior as well as the asymmetry production via analytical approximations.
\begin{itemize}
\item Annihilations drop out of equilibrium at $z_f$ when $\Gamma_{\rm ann}^S(z_f)\simeq H(z_f)$. This freeze-out temperature depends on the annihilation cross section only logarithmically 
\begin{align}
z_f\sim \ln\left[5\times 10^7\left(\frac{M}{300~{\rm GeV}}\right)\left(\frac{\sigma_0}{1~{\rm fb}}\right)\right], \label{eq:zf}
\end{align}
which gives $z_f\sim 11-18$ for $\sigma_0={\rm ab-fb}$.

If there were no decays, this would give the freeze-out density of particles, as in the WIMP case,
\begin{align}
\Sigma(z_f)\sim 10^{-10}\left(\frac{300~{\rm GeV}}{M}\right)\left(\frac{1~{\rm fb}}{\sigma_0}\right) z_f^2. \label{eq:sigmazf}
\end{align}
However in our case the remaining particles decay with a decay rate $\Gamma$. We can approximate the total number density at later times as
\begin{align}
\Sigma(z)\simeq C\exp\left(-\frac{\Gamma}{2H(z)}\right)\Sigma(z_f)\quad \text{for}\quad z>z_f,
\end{align}
where $C$ is a numerical factor that can be found by matching to $\Sigma_{\rm eq}(z)=\frac{45}{\pi^2\sqrt{2\pi}g_\ast}z^{3/2}e^{-z}$ at $z\simeq z_f$.

\item Flavor-blind annihilations cause decoherence, as can be seen from the equations for $\Pi(z)$ and $\Xi(z)$.  Due to this decoherence, oscillations are further delayed. In order to (approximately) find when oscillations start in the presence of flavor-blind annihilations we look at the Boltzmann equations for $\Delta(z)$ and $\Xi(z)$, setting $\Gamma=0$. (Decays are important for $CP$-violation, but less so for oscillations themselves.) We then arrive at an equation for a damped harmonic oscillator, similar to the one in Eq.~\ref{eq:Delta}:
\begin{align}
\frac{d^2\Delta(y)}{dy^2}+2\,\xi\,\omega_0\,\frac{d\Delta(y)}{dy}+\omega_0^2\,\Delta(y)=0, \label{eq:Deltaann}
\end{align}
where $y=z^2$ and 
\begin{align*}
\omega_0\equiv \frac{m}{yH},\quad \xi\equiv \frac{\Gamma_{\rm ann}^S}{2m},
\end{align*}
with the identification $2\,\Gamma_{\rm ann}^S=s\langle \sigma_{\rm ann}^S v\rangle\Sigma(z)$. This equation cannot be solved analytically since $\xi$ is a function of $z$. Specifically $\xi$ decreases with decreasing temperature/growing $z$. For early times, $\xi\gg1$, the system is overdamped and there are no oscillations. Oscillations only start when $\xi<1$,  $\omega_{\rm osc}\sim \Gamma_{\rm ann}^S(z_{\rm osc})$, which gives
\begin{align}
z_{\rm osc}\sim& \ln\left[10^{7}\left(\frac{M}{300~{\rm GeV}}\right)^3\left(\frac{2\times 10^{-6}~{\rm eV}}{m}\right)\left(\frac{\sigma_0}{1~{\rm fb}}\right)\right]. \label{eq:zoscann}
\end{align}
For example for $\sigma_0=1~$fb oscillations start at $z_{\rm osc}\sim 16$. (See Fig.~\ref{fig:zosc}.) 

\item The $\psi$-asymmetry $\Delta(z)$ is also suppressed due to flavor-blind annihilations. We show this asymmetry in Fig.~\ref{fig:sigmadelta} for different annihilation cross sections. In order to find an approximate expression for $\Delta(z)$, first realize that $\Gamma\sim m\sim \Gamma_{\rm ann}$ when the oscillations start. (We take $\Gamma\sim m$ to maximize the $CP$-violation in oscillations.) Hence, immediately following the start of oscillations $\Gamma_{\rm ann}\ll \Gamma$ and we can ignore it in the Boltzmann equations. Furthermore oscillations start before annihilations freeze out and $\psi$-particles oscillate a few times before they decay for interaction scales we consider ($\sigma_0={\rm ab-fb}$). Then we can solve Eq.~\ref{eq:Delta} for $z>z_{\rm osc}$ with 
\begin{align*}
\Sigma(z)\simeq 2\,Y_{\rm eq}(z_{\rm osc}) \exp\left( -\frac{\Gamma}{2H(z)}\right),
\end{align*}
and with symmetric initial conditions to find the $\psi$-asymmetry
\begin{align}
\Delta (z)\simeq  \epsilon\,Y_{\rm eq}(z_{\rm osc}) \exp\left( -\frac{\Gamma}{2H(z)}\right)\sin^2\left(\frac{m}{2H(z)}\right). \label{eq:Deltablindapp}
\end{align}

We emphasize that the behavior before the oscillations start, where the annihilation rate is much larger than the mass difference and the decay rate, is not covered in this approximation. We also ignore annihilations altogether right after the oscillations start. However there is a window where $\omega_{\rm osc}>\Gamma_{\rm ann}\gtrsim \Gamma$ for $z>z_{\rm osc}$, which should affect the size of the asymmetry as well as the frequency of the oscillations. Furthermore we omitted the freeze out of annihilations. Hence the above approximation is expected to underestimate the asymmetry. Still it estimates the maximum $\psi$-asymmetry within an order of magnitude for the  parameters given in Eq.~\ref{eq:parameters}. In Fig.~\ref{fig:deltaapp} we compare this approximation to the numerical solutions of Eq.~\ref{eq:Bflavorblind}.
\end{itemize}

\subsection{Flavor-Sensitive Interactions} \label{sec:flavor-sensitive}
\begin{figure}[h!]
\includegraphics[width=\linewidth]{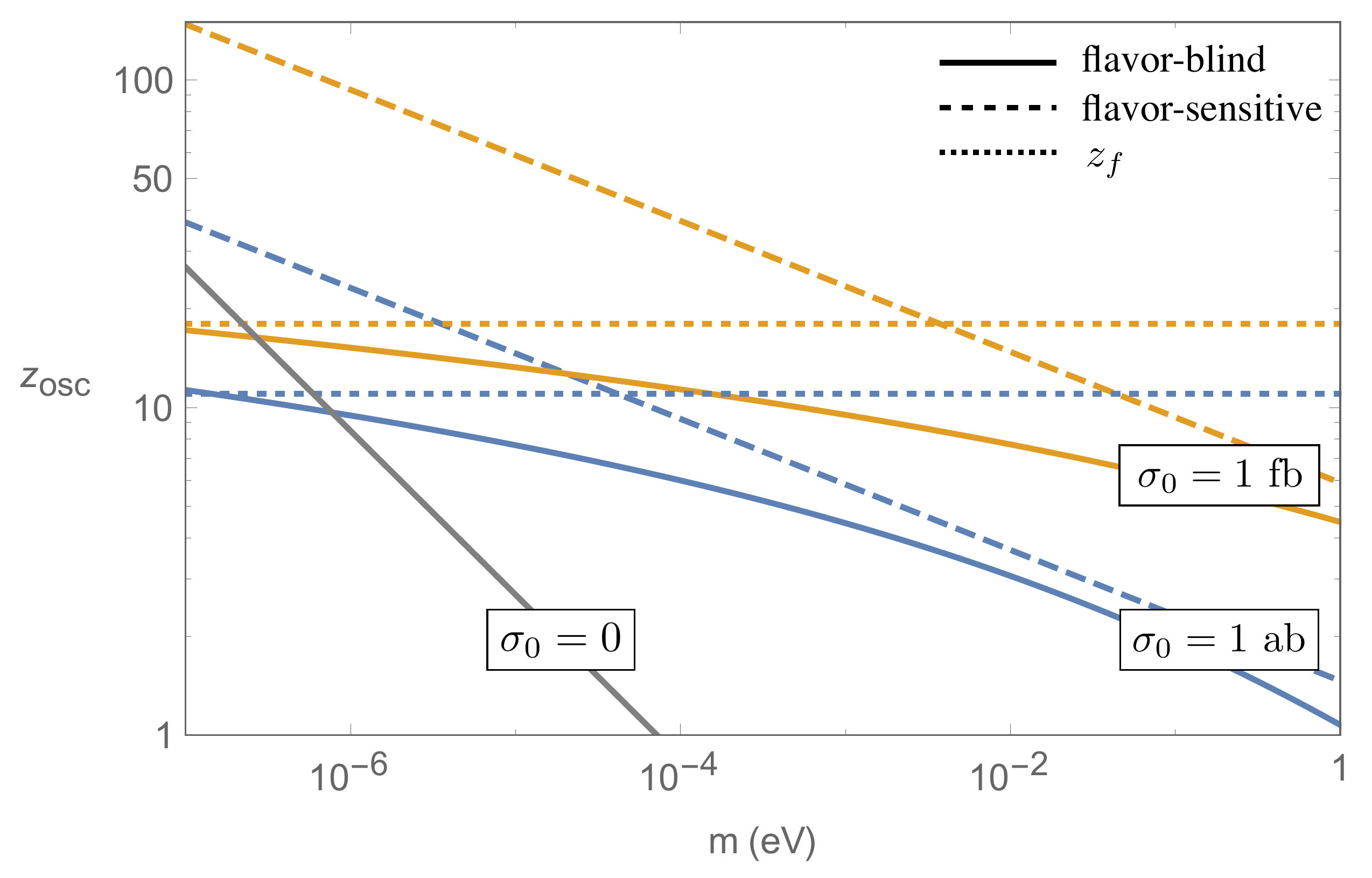}
\caption{The time when the oscillations start, $z_{\rm osc}$, vs the Majorana mass $m$ for flavor-blind (solid, blue/orange) and flavor-sensitive (dashed, blue/orange) interactions as well as  the Hubble suppression without interactions (solid, gray). (Plotted are Eqs.~\ref{eq:zoscH}, \ref{eq:zoscann} and  \ref{eq:zoscscat}.) Note that the oscillation frequency is related to the Majorana mass as $\omega_{\rm osc}=2m$. We also show the freeze-out temperature $z_f$ (dotted) given in Eq.~\ref{eq:zf}.}\label{fig:zosc}
  \end{figure}

Now let us investigate the effects of vector interactions, which are flavor sensitive. Particularly important in this case are elastic scatterings. If a scattering process probes the particle or antiparticle nature of the $\psi$-particles, oscillations cannot proceed. (This is called the \emph{quantum Zeno effect}.) This can be seen from the Boltzmann equations with flavor-sensitive interactions:
\begin{widetext}
\begin{align}
\frac{d\Delta(z)}{dz}&=-\frac{\Gamma}{zH(z)}\,\Delta(z)+i\frac{2m}{zH(z)}\,\Xi(z) ,\notag\\
\frac{d\Sigma(z)}{dz}&=-\frac{\Gamma}{zH(z)}\,\Sigma(z)-\frac{2\,\Gamma\,r\cos\phi_\Gamma}{zH(z)}\Pi(z)+i\frac{2\Gamma\,r\sin\phi_\Gamma}{zH(z)}\Xi(z) -\frac{s(z)\langle \sigma_{\rm ann}^V v\rangle(z)}{2zH(z)}\biggl(\Sigma^2(z)-\Delta^2(z)+\Xi^2(z)-\Pi^2(z)-4Y_{\rm eq}^2(z)\biggr), \notag \\
\frac{d\Xi(z)}{dz}&=-\frac{\Gamma+2\,\Gamma_{\rm scat}^V(z)}{zH(z)}\,\Xi(z)+i\frac{2m}{zH(z)}\,\Delta(z)-i\frac{2\Gamma\,r\sin\phi_\Gamma}{zH(z)} \Sigma(z), \notag\\
\frac{d\Pi(z)}{dz}&=-\frac{\Gamma+2\,\Gamma_{\rm scat}^V(z)}{zH(z)}\,\Pi(z)-\frac{2\,\Gamma\,r\cos\phi_\Gamma}{zH(z)}\Sigma(z). \label{eq:Boltsens}
\end{align}
\end{widetext}
Following the arguments of the previous section to describe the $\psi$-asymmetry, one can show that oscillations start only when $\omega_{\rm osc}\sim \Gamma_{\rm scat}^V(z_{\rm osc})$, \emph{i.e.}
\begin{align}
z_{\rm osc}\simeq 80 \left(\frac{M}{300~{\rm GeV}}\right)^{3/5}\left(\frac{2\times 10^{-6}~{\rm eV}}{m}\right)^{1/5}\left(\frac{\sigma_0}{1~{\rm fb}}\right)^{1/5}. \label{eq:zoscscat}
\end{align}
Compared to the flavor-blind annihilations, oscillations are delayed much further due to flavor-sensitive scatterings (with similar cross sections). This is expected since elastic scatterings off light particles in the plasma are not Boltzmann suppressed at temperatures $T\sim O(100~{\rm GeV})$. Oscillations start at $z\sim 80$ for a flavor-sensitive elastic scattering cross section $\sigma_0=1~$fb (compared to $z\sim 18$ for flavor-blind annihilations.) Since the $\psi$-particle number density is already less than $10^{-10}$ by $z\sim 40$, the asymmetry produced (after the oscillations start) would be too small compared to the BAU. In Fig.~\ref{fig:zosc} we show $z_{\rm osc}$ vs the mass difference for different interaction strengths.  

\begin{figure*}[!htb]
\includegraphics[width=\textwidth]{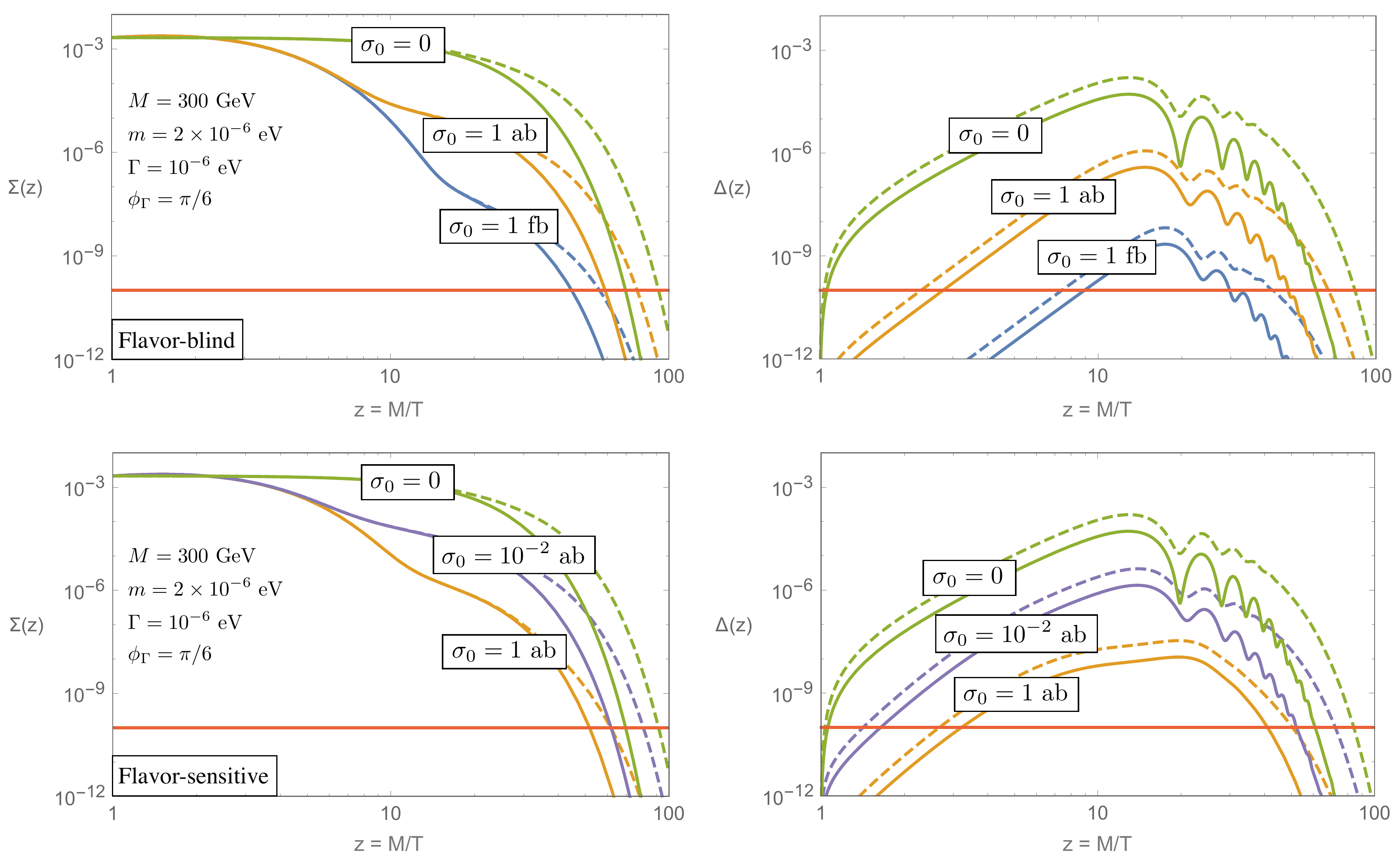}
\caption{\textbf{(Left)}The total $\psi$-number density $\Sigma(z)$ and \textbf{(Right)} the $\psi$-asymmetry $\Delta(z)$ (with symmetric initial conditions) for different cross-sections. The following parameters are used in both plots: $M=300$~GeV, $\Gamma=10^{-6}$~eV, $m=2\times 10^{-6}$~eV, $r=0.1$ (solid), 0.3 (dashed), $\sin\phi_\Gamma=0.5$. \textbf{(Top)} Flavor-blind interactions delay oscillations due to annihilations, until $\omega_{\rm osc}\sim \Gamma_{\rm ann}$. This corresponds to $z_{\rm osc}\sim 16~(9)$ for an annihilation cross section of 1~fb (1~ab). \textbf{(Bottom)} Flavor-sensitive interactions delay oscillations due to elastic scatterings, until $\omega_{\rm osc}\sim \Gamma_{\rm scat}$. This corresponds to $z_{\rm osc}\sim 20~(8)$ for an annihilation cross section of 1~ab ($10^{-2}$~ab). Without interactions the oscillations start at $z_{\rm osc}\sim 6$. $CP$ violation, and hence the asymmetry, is smaller for larger cross sections since oscillations are delayed longer. The baryon asymmetry of the Universe, $\eta\simeq 10^{-10}$, is shown for reference.  }\label{fig:sigmadelta}
  \end{figure*}

An approximation for the $\psi$-asymmetry  can be found following the steps that led to Eq.~\ref{eq:Deltablindapp}, with one change. Now the oscillations start after annihilations freeze out. Hence we solve Eq.~\ref{eq:Delta} for $z>z_{\rm osc}$ with 
\begin{align*}
\Sigma(z)\simeq \exp\left(-\frac{\Gamma}{2H(z)}\right)\Sigma(z_f),
\end{align*}
and find the asymmetry
\begin{align}
\Delta (z)\simeq  \frac{\epsilon\, \Sigma(z_f)}{2} \exp\left( -\frac{\Gamma}{2H(z)}\right)\sin^2\left(\frac{m}{2H(z)}\right), \label{eq:Deltasensapp}
\end{align}
where $\Sigma(z_f)$ is given in Eq.~\ref{eq:sigmazf}. We show the comparison between the above approximation and the numerical results in Fig.~\ref{fig:deltaapp}. We also show the numerical solutions to Eq.~\ref{eq:Boltsens} for $\Sigma(z)$ and $\Delta(z)$ in Fig.~\ref{fig:sigmadelta}.

\section{(Approximate) Baryon Asymmetry of the Universe}\label{sec:BAU}
\subsection{Baryon asymmetry via $B$-violating interactions}
\begin{figure*}
\includegraphics[width=\linewidth]{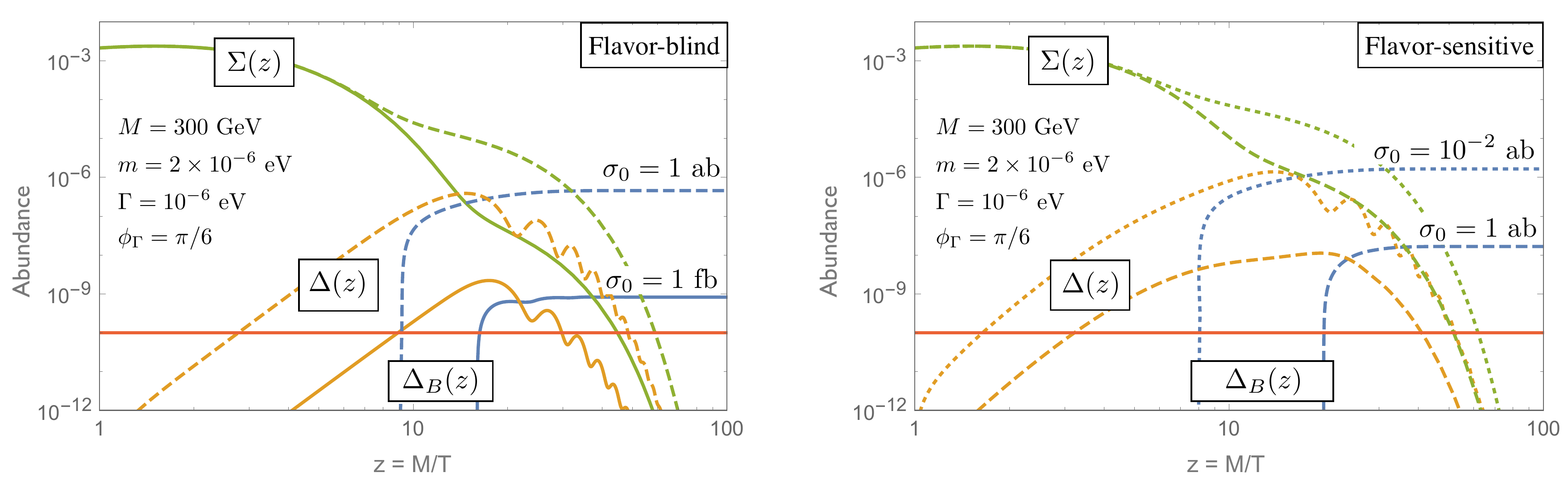}
\caption{The total $\psi$-number density $\Sigma(z)$, the $\psi$-asymmetry $\Delta(z)$ and the baryon asymmetry $\Delta_B(z)$  for $\sigma_0=1~$fb (solid), 1~ab (dashed), $10^{-2}~$ab (dotted) and $M=300~{\rm GeV},~ m=2\times 10^{-6}~{\rm eV},~\Gamma=10^{-6}~{\rm eV},~r=0.1,~\sin\phi_\Gamma=0.5$. The baryon asymmetry of the Universe, $\eta\simeq 10^{-10}$, is shown for reference. The oscillations are delayed longer for flavor-sensitive interactions: For an effective cross section $\sigma_0=1~$ab (dashed) the oscillations start at $z_{\rm osc}\sim9$ if the interaction is flavor blind, while they start at $z_{\rm osc}\sim20$ if the interaction is flavor sensitive. With the parameters used, not enough baryon asymmetry is produced for $\sigma_0\gtrsim10~$ab with flavor-sensitive interactions}\label{fig:basym}
\end{figure*}

So far we have only discussed the $\psi$-asymmetry. However, our main purpose is to produce the baryon asymmetry of the Universe. For that we need to add to the set of Boltzmann equations in Eq.~\ref{eq:oscBoltzmann} two more equations describing the evolution of baryon and antibaryon densities.  These equations take into account processes that change  the baryon number, such as inelastic scatterings $BX_1\to \psi X_2$. Obviously one needs to know the details of the model to properly set up and solve the relevant Boltzmann equations. We will do this after we introduce a model in Section~\ref{sec:BAUreal}. However, as long as the oscillations are delayed till $z> 1$, the general workings of this scenario are quite robust towards the details of a model and it is helpful to give an approximate picture. Hence we first focus on  baryon-number-violating terms in Eq.~\ref{eq:Lint} and assume that there are no other baryon-number-violating interactions. (Baryogenesis via the lepton-number-violating terms is relatively straightforward and we will mention it in the next section.) 

Before solving for the baryon asymmetry, let us revisit the oscillation dynamics in the presence of flavor-blind annihilations as they relate to the production of a baryon asymmetry. (Flavor-sensitive interactions follow a similar story.) 

For a mass difference of $2\times 10^{-6}~$eV and effective cross-section $\sigma_0={\rm ab-fb}$, oscillations start at $z_{\rm osc}\sim9-16$. At this point the annihilation rate is Boltzmann suppressed and drops below the decay rate very quickly. The Hubble rate is already much smaller than the mass difference for $z>6$ (see Fig.~\ref{fig:ratesvz}). Hence when the oscillations start they proceed as described in Section~\ref{sec:osc}. Furthermore $\psi$-particles oscillate a few times before they decay. (Note that this is very different from soft leptogenesis models in which oscillations are thought to start at $z\lesssim1$.)

With these in mind we can write the Boltzmann equations for the baryon and antibaryon number densities for $z>z_{\rm osc}\gg1$ as
\begin{align}
\frac{dY_B}{dz}&=\frac{\Gamma_\psi}{zH(z)}\, Y_\psi+\frac{\Gamma_{\psi^c}}{zH(z)}\, Y_{\psi^c}, \notag \\
\frac{dY_{\bar{B}}}{dz}&=\frac{\overline{\Gamma}_\psi}{zH(z)}\, Y_\psi+\frac{\overline{\Gamma}_{\psi^c}}{zH(z)} \,Y_{\psi^c},
\end{align}
where $\Gamma_\psi\equiv\Gamma(\psi \to BX),~ \overline{\Gamma}_\psi\equiv\Gamma(\psi \to \bar{B}\, \bar{X})$ (and similarly for $\psi^c$). We ignore inverse decays $B\to X\psi$ for $T\ll M$. We make the following approximations for $z>z_{\rm osc}$,
\begin{align*}
&\Gamma_\psi+\overline{\Gamma}_\psi \simeq \Gamma_{\psi^c}+\overline{\Gamma}_{\psi^c}\simeq \Gamma, \\
&\Gamma_\psi-\overline{\Gamma}_\psi \simeq \Gamma_{\psi^c}-\overline{\Gamma}_{\psi^c}\simeq \epsilon\,\Gamma.
\end{align*}
Defining 
\begin{align*}
\Sigma_B= Y_B+Y_{\bar{B}}~,\quad \quad \Delta_B=Y_B-Y_{\bar{B}}~,
\end{align*}
the differential equations for the total baryon number and the baryon asymmetry are
\begin{align}
\frac{d\Sigma_B(z)}{dz}=\frac{\Gamma}{zH(z)}\,\Sigma(z), \quad 
 \frac{d\Delta_B(z)}{dz}=\frac{\epsilon\,\Gamma}{zH(z)}\,\Sigma(z). \label{eq:asymmetry}
\end{align}
The baryon asymmetry is proportional to the $CP$ violation, parametrized by $\epsilon$. These equations are solved together with Eq.~\ref{eq:oscBoltzmann}. Corresponding baryon asymmetries are shown in Fig.~\ref{fig:basym} in the presence of flavor-blind or flavor-sensitive interactions. A large enough baryon asymmetry can be produced with flavor-blind interactions with cross sections as high as $O$(fb). The delay of oscillations is stronger for flavor-sensitive interactions. In this case, in order to produce an asymmetry of $10^{-10}$, the elastic scattering cross section should be $O$(ab) or less.

Let us emphasize that we assume that a nonzero baryon asymmetry is only produced after the oscillations start, setting $\Delta(z<z_{\rm osc})=0$. However, even though oscillations are suppressed for $z<z_{\rm osc}$, some $CP$-asymmetry is produced. ( This can be seen as a nonzero $\psi$-asymmetry for $z<z_{\rm osc}$ in Fig.~\ref{fig:sigmadelta}.) In either case, the maximum asymmetry is approximated well by
\begin{align}
\Delta_B(z>z_{\rm osc})\simeq \epsilon\,\Sigma(z_{\rm osc}).
\end{align}

\begin{figure}
\includegraphics[width=\linewidth]{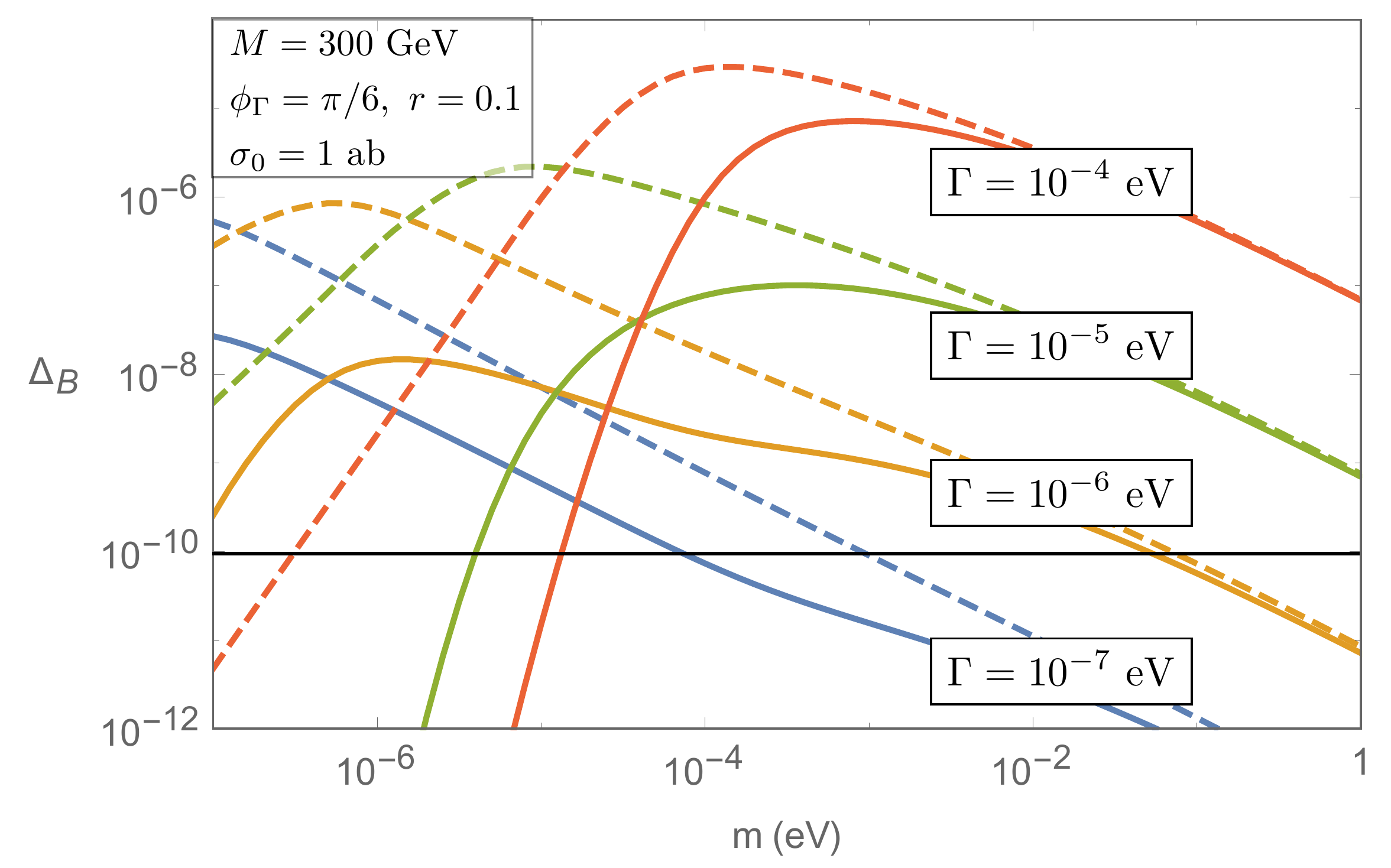}
\caption{The final baryon asymmetry, $\Delta_B(z\to\infty)$ vs the Majorana mass for different decay rates and for both flavor-blind (dashed) and flavor-sensitive (solid) interactions with $\sigma_0=1~$ab, $M=300~$GeV, $r=0.1$, and $\sin\phi_\Gamma=0.5$. For mass differences ($\Delta m=2m$) larger than $\sim\!10^{-2}$~eV, oscillations start $z_{\rm osc}<5$ and there is little difference between flavor-blind and -sensitive interactions. If the mass difference is smaller than $10^{-2}~$eV, oscillations are delayed longer for flavor-sensitive interactions. (See also Fig.~\ref{fig:zosc}.) Hence the baryon asymmetry is smaller (compared to flavor-blind interactions).}\label{fig:DBvm}
\end{figure}

Note that for larger mass differences the oscillations start earlier, as can be seen in Fig.~\ref{fig:zosc}. Since $\Sigma(z_{\rm osc})$ is larger for smaller $z_{\rm osc}$, one might expect to get a larger baryon asymmetry for a larger mass difference. However for a given decay rate, as the mass difference gets larger, $CP$ violation ($\epsilon\propto \Gamma/m$) gets smaller. In Fig.~\ref{fig:DBvm} we show the final baryon asymmetry, $\Delta_B(z\to\infty)$, for different mass differences and decay rates for both flavor-blind and flavor-sensitive interactions with $\sigma_0=1~$ab. There is a wide range of decay rates and mass differences that can accommodate the correct baryon asymmetry of the Universe.

\subsection{Baryon asymmetry via $L$-violating interactions}
In the previous sections we focused on baryon-number-violating interactions. If, however, lepton-number-violating terms in Eq.~\ref{eq:Lint} dominate, the picture slightly changes. The lepton--antilepton asymmetry is still given by Eq.~\ref{eq:asymmetry} (by just changing the label $B\to L$) under similar assumptions. If we ignore the baryon-number-violating terms, a $(B-L)$-asymmetry is produced in this case. If this asymmetry is produced before the electroweak transition, it can be turned into a baryon-asymmetry by sphalerons, which are active during the EW transition, $T\simeq130~$GeV~\cite{D'Onofrio:2014kta}. The baryon asymmetry produced by $(B-L)$-conserving sphaleron processes is given by (for $M=300~$GeV)
\begin{align}
\Delta_B =-\frac{22+4n_H}{66+13n_H} \Delta_L(z\sim 2),
\end{align}
where $n_H$ is the number of Higgs doublets. 

A few remarks are in order at this point. In order to produce a lepton asymmetry before the EW transition, oscillations should start at $T\gtrsim 130~$GeV. This means that the oscillation frequency $\omega_{\rm osc}> H(T\sim 130~{\rm GeV})\simeq 10^{-5}~$eV.  With $O$(ab) annihilation cross sections and with $\Gamma\sim m\gtrsim 10^{-5}$ it is possible to produce enough lepton asymmetry before the EW transition. However, note that with the parameters used in Eq.~\ref{eq:parameters}, the oscillations start at $T\ll100~$GeV even without any annihilations. Hence not enough lepton asymmetry is produced before the EW transition for the benchmark scenario.

\section{Baryogenesis via Pseduo-Dirac Bino Oscillations} \label{sec:BAUreal}
The scenario described in the previous sections can be realized in any UV theory with pseudo-Dirac fermions. In this section, as a concrete example, we show that pseudo-Dirac bino oscillations within the model introduced in Ref.~\cite{Ipek:2014moa}\footnote{In Ref.~\cite{Ipek:2014moa} the focus was gluino interactions. Gluinos interact strongly. Their annihilation cross section would be too big to fall out of equilibrium. Hence we study bino interactions here.} could generate the baryon asymmetry of the Universe. 

\subsection{The Model}\label{sec:model}

The model we study is a SUSY model with an approximate global $U(1)_R$ symmetry. The SM particles are not charged under this global $U(1)_R$ while all the supersymmetric partners have +1 $R$-charge. With this $R$-charge assignment, the gauginos cannot have Majorana masses. In order to give Dirac mass to the bino we introduce the super field $\Phi_S$ whose fermion component $S$, the singlino, is the Dirac partner of the bino. In order to give nongauge couplings to the singlino, we introduce the superfields $\Phi_{\bar{D}}$ and $\Phi_D$, transforming under the SM gauge group in the same way as $\bar{d}$ and $\bar{d}^\ast$, respectively.  The field content of the model that is relevant for us is shown in Table~\ref{table:fields}. We will only give a short summary of the complete model focusing on the parts that are most relevant to baryogenesis. For details see Ref.~\cite{Ipek:2014moa}. 

\begin{table} [h]
\begin{tabular}{|c|c|c|c|c|}
\hline
Fields	&	~$SU(3)_c$~	&	~$SU(2)_L$~	&	~$U(1)_Y$~	&	~$U(1)_R$~ \\
\hline\hline
$Q=\tilde{q}+\theta\, q$	&	3	&	2	&	1/6	&	1 \\
$\bar{U}=\sfermion{u}+\theta\, \bar{u}$	&	$\bar{3}$		&	1	&	-2/3	&	1	\\
$\bar{D}=\sfermion{d}+\theta\, \bar{d}$	&	$\bar{3}$		&	1	&	1/3	&	1	\\	
$L=\tilde{\ell}+\theta\, \ell$	&	1	&	2	&	-1/2	&	1 \\
$\bar{E}=\sfermion{e}+\theta\, \bar{e}$	&	$1$		&	1	&	1	&	1	\\	
$\Phi_{\bar{D}}=\phi_{\bar{D}}+\theta\,\psi_{\bar{D}}$	&	$\bar{3}$		&	1	&	1/3	&	1	\\
$\Phi_{D}=\phi_{D}+\theta\,\psi_{D}$	&	3		&	1	&	-1/3	&	1	\\
\hline
$W_{\tilde{B},\alpha}\supset\tilde{B}_\alpha$	&	1	&	1	&	0	&	1	\\
$\Phi_S=\phi_s+\theta\, S$	&	1	&	1	&	0	&	0	\\
\hline

\end{tabular}
\caption{Part of the particle content and their associated quantum numbers under the SM gauge group and $U(1)_R$.  All the fermion fields are left-handed Weyl spinors. $\tilde{B}$ is the bino and $S$ is its Dirac partner, the singlino.  $\phi_D$, $\phi_{\bar{D}}$ are superpartners of some exotic heavy vectorlike quarks. $q,\bar{u},\bar{d},\ell,\bar{e}$ are the SM fermion fields. Generational indices are suppressed for simplicity.} \label{table:fields}
\end{table}

The mass Lagrangian for the bino and the singlino is
\begin{align}
-\mathcal{L}_{\rm mass} &= M_D \tilde{B}S+\frac12\left(m_{\tilde{B}}\tilde{B}\tilde{B}+m_S SS \right) + {\rm h.c.},
\end{align}
where $M_D$ is the Dirac mass and $m_{\tilde{B},S}$ are $U(1)_R$-breaking Majorana masses. The Dirac mass
\begin{align}
M_D=\frac{c D}{\Lambda_M},
\end{align}
arises from a spurion term where $c$ is a dimensionless parameter, $D$ is a SUSY-breaking order parameter and $\Lambda_M$ is the messenger scale. Majorana mass terms for the gauginos will be generated by anomaly mediation \cite{Randall:1998uk, Giudice:1998xp, ArkaniHamed:2004yi}, which gives, \emph{e.g.}, a Majorana bino mass
\begin{align}
m_{\tilde{B}}=\frac{\beta(g_Y)}{g_Y}F_\phi.
\end{align}
$\beta(g_Y)$ is the beta function for the hypercharge coupling constant $g_Y$ and $F_\phi$ is a conformal parameter satisfying
\begin{align}
\frac{m_{3/2}^3}{16\pi^2 M_{\rm Pl}^2}\lesssim |F_\phi| \lesssim m_{3/2}.
\end{align}
$m_{3/2}$ is the gravitino mass. Note that we do not need  a light gravitino to have a small Majorana mass for the bino. We assume that the gravitino is heavy enough (heavier than $\sim\!{\rm keV}$) such that binos mostly decay to SM fermions (via R-parity-violating interactions). A Majorana mass for the singlino could arise from the $U(1)_R$-violating superpotential term
\begin{align}
\int d^2\theta\, m_S \Phi_S^2 + {\rm h.c.}
\end{align}
We assume all $U(1)_R$-violating terms are small, $m_S\ll M_D$. Then we can define the \emph{pseudo-Dirac bino} in this model as
\begin{align}
\psi_{\tilde{B}}=\left(\begin{array}{c} \tilde{B} \\
					S^\dagger\end{array}\right),
\end{align}
and follow the oscillation picture described in Section~\ref{sec:osc} where bino--antibino states mix. (It should be clear from context if the word ``bino" refers to the Weyl spinor $\tilde{B}$ or the pseudo-Dirac fermion $\psi_{\tilde{B}}$.) We take the lightest neutralino to be purely bino so that there is no mixing between, for example, the bino, singlino and the Dirac partner of the wino. 

$U(1)_R$-conserving interactions of the bino and the singlino include
\begin{align}
-\mathcal{L}~\supset~ \sqrt{2}\,g_Y Y_R\,\tilde{B}\, \bar{d}_i \tilde{\bar{d}}_i +y_i\, s\,\bar{d}_i  \phi_D + {\rm h.c.},\label{eq:muDgS}
\end{align}
where $Y_R$ is the hypercharge of the right-handed down-type quark.
 
The new scalars $\phi_D,\phi_{\bar{D}}$ can be assumed to be degenerate with mass $\mu_D$ and with the mass mixing term
\begin{align}
B_{D\bar{D}}^2\phi_D\phi_{\bar{D}} + {\rm h.c.},
\end{align}
where $B_{D\bar{D}}^2=\frac{c_{D\bar{D}}\,D^2}{\Lambda_M^2}$ and $c_{D\bar{D}}$ a constant.

\subsubsection{R-Parity Violating Bino Decays}
In order to have $CP$ violation in pseudo-Dirac bino oscillations, the bino must decay. We assume that the bino is the lightest $R$-charged particle and decays via $U(1)_R$-breaking interactions. We also assume R parity is broken so that there is baryon/lepton number violation. (For an extended review of R-parity-violating interactions and phenomenological constraints, see, for example, \cite{Barbier:2004ez}.) We include the following R-parity- and $U(1)_R$-symmetry-violating interactions
\begin{align}
W_{\slashed{L}} &= \lambda_{ijk}' L_iQ_j \bar{D}_k + \lambda_{ij}' L_i Q_j \Phi_{\bar{D}} +{\rm h.c.},\notag\\
W_{\slashed{B}}&=\frac12 \lambda_{ijk}'' \bar{U}_i\bar{D}_j\bar{D}_k +\lambda_{ij}'' \bar{U}_i\bar{D}_j \Phi_{\bar{D}}+{\rm h.c.}
\end{align}
$W_{\slashed{L}}$ has lepton-number-violating terms, while $W_{\slashed{B}}$ has baryon-number-violating terms. The supersymmetric Lagrangian contains the interactions
\begin{align}
-\mathcal{L}~\supset~& \lambda_{ijk}'\ell_i q_j \tilde{\bar{d}}_k + \lambda_{ij}' \ell_i q_j \phi_{\bar{D}}\notag\\ 
&+ \frac12\lambda_{ijk}'' \bar{u}_i\bar{d}_j\tilde{\bar{d}}_k+\lambda_{ij}''\bar{u}_i\bar{d}_j\phi_{\bar{D}}+{\rm h.c.} \label{eq:RPV}
\end{align}

Let us now assume that all the squarks and $\phi_D,\phi_{\bar{D}}$ are heavier than the bino and can be integrated out. Then the effective four-fermion Lagrangian is
\begin{align}
-\mathcal{L}_{\rm eff}=&g'_{\tilde{B},ijk}\tilde{B}\ell_i q_j\bar{d}_k+g'_{S,ijk}S\ell_i q_j\bar{d}_k \notag\\
&+g''_{\tilde{B},ijk}\tilde{B}\bar{u}_i\bar{d}_j\bar{d}_k+g''_{S,ijk}S\bar{u}_i\bar{d}_j\bar{d}_k +{\rm h.c.}, \label{eq:Leff}
\end{align}
with
\begin{align}
&g'_{\tilde{B},ijk}=\frac{\sqrt{2}\,g_Y\lambda'_{ijk}}{3\,m_{\rm sf}^2},\quad\quad\quad g'_{S,ijk}=\frac{y_k\,\lambda'_{ij}B_{D\bar{D}}^2}{\mu_D^4},\\
&g''_{\tilde{B},ijk}=\frac{\sqrt{2}\,g_Y\lambda''_{ijk}}{3\,m_{\rm sf}^2}, \quad\quad\quad g''_{S,ijk}=\frac{y_k\, \lambda''_{ij}B_{D\bar{D}}^2}{\mu_D^4},\notag
\end{align}
where $m_{\rm sf}$ is a common sfermion mass. We assume that $\phi_D,\phi_{\bar{D}}$ are heavier than the squarks such that $|g_{\tilde{B}}'|\gg |g_S'|$ and $|g_{\tilde{B}}''|\gg |g_S''|$.

Comparing Eq.~\ref{eq:Leff} with Eq.~\ref{eq:Lint} and assuming one generation of fermions we can identify
\begin{align}
g_\chi&\equiv g_{\tilde{B}}'',\quad g_\chi'\equiv g_{\tilde{B}}', \qquad
g_\eta\equiv g_S'',\quad g_\eta'\equiv g_S'.
\end{align}
If the baryon-number-violating terms dominate over the lepton-number-violating ones, then the decay rate is
\begin{align}
\Gamma&=\frac{M_D^5}{(32\pi)^3}\left(|g_{\tilde{B}}''|^2+|g_S''|^2\right).
\end{align}
For $|g_{\tilde{B}}''|\gg |g_S''|$, the decay rate can be parametrized as
\begin{align}
\Gamma\simeq 10^{-6}~{\rm eV}\left(\frac{M_D}{300~{\rm GeV}}\right)^5\left(\frac{10~{\rm TeV}}{m_{\rm sf}}\right)^4\left(\frac{\lambda''}{10^{-2}}\right)^2.
\end{align}

\subsubsection{Bino Annihilations and Elastic Scatterings}
As discussed in Section~\ref{sec:oscearly} annihilations and elastic scatterings of the binos are very important in studying pseudo-Dirac bino oscillations in the early Universe. For small mass splittings ($m\ll~$eV), we can treat the binos as purely Dirac to find the annihilation and elastic scattering cross sections. Since we assume that the lightest neutralino is a pure bino, even for binos heavier than $W/Z$ bosons, annihilations into fermion final states dominate~\cite{Griest:1989zh}. Hence we use the effective Lagrangian
\begin{align}
-\mathcal{L}_{\rm scat}=\frac{g_Y^2}{m_{\rm sf}^2}\bar{\psi}_{\tilde{B}}\gamma_\mu P_L\psi_{\tilde{B}}\,\bar{F}\gamma^\mu(g_V+g_A\gamma_5)F,
\end{align}
where 
\begin{align*}
g_V=\frac{Y_R^2+Y_L^2}{2}, \quad g_A=\frac{Y_R^2-Y_L^2}{2}, \quad F = \left(\begin{array}{c} f_L \\
															f_R^\dagger \end{array}\right),
\end{align*}
and $Y_{L,R}$ is the hypercharge of the fermion $f_{L,R}$. The thermally averaged annihilation cross section is~ \cite{Hsieh:2007wq}
\begin{align}
\langle\sigma_{\rm ann}v\rangle = \frac{g_Y^4 M_D^2}{8\pi m_{\rm sf}^4}\sum_{f} N_f Y_f^4\simeq 0.6\, \frac{g_Y^4 M_D^2}{\pi m_{\rm sf}^4},
\end{align}
where the sum is over all SM fermions. The color factor $N_f=3$ for quarks and 1 for leptons. 

The bino has both flavor-blind (axial-vector) and flavor-sensitive (vector) interactions. Since the annihilation rate is exponentially suppressed compared to the elastic scattering rate, the delay in oscillations is governed by the flavor-sensitive scattering part of the interactions. Then the relevant thermally-averaged scattering cross-section is
\begin{align}
\langle\sigma_{\rm scat}v\rangle=\frac{g_Y^4T^2}{16\pi m_{\rm sf}^4}\sum_f N_f\, Y_f^4\simeq 0.3\, \frac{g_Y^4T^2}{\pi m_{\rm sf}^4},
\end{align}
where the sum is taken over all SM fermions except the top quark.

\subsection{Baryon Asymmetry of the Universe} 

\begin{figure}[h!]
\includegraphics[width=\linewidth]{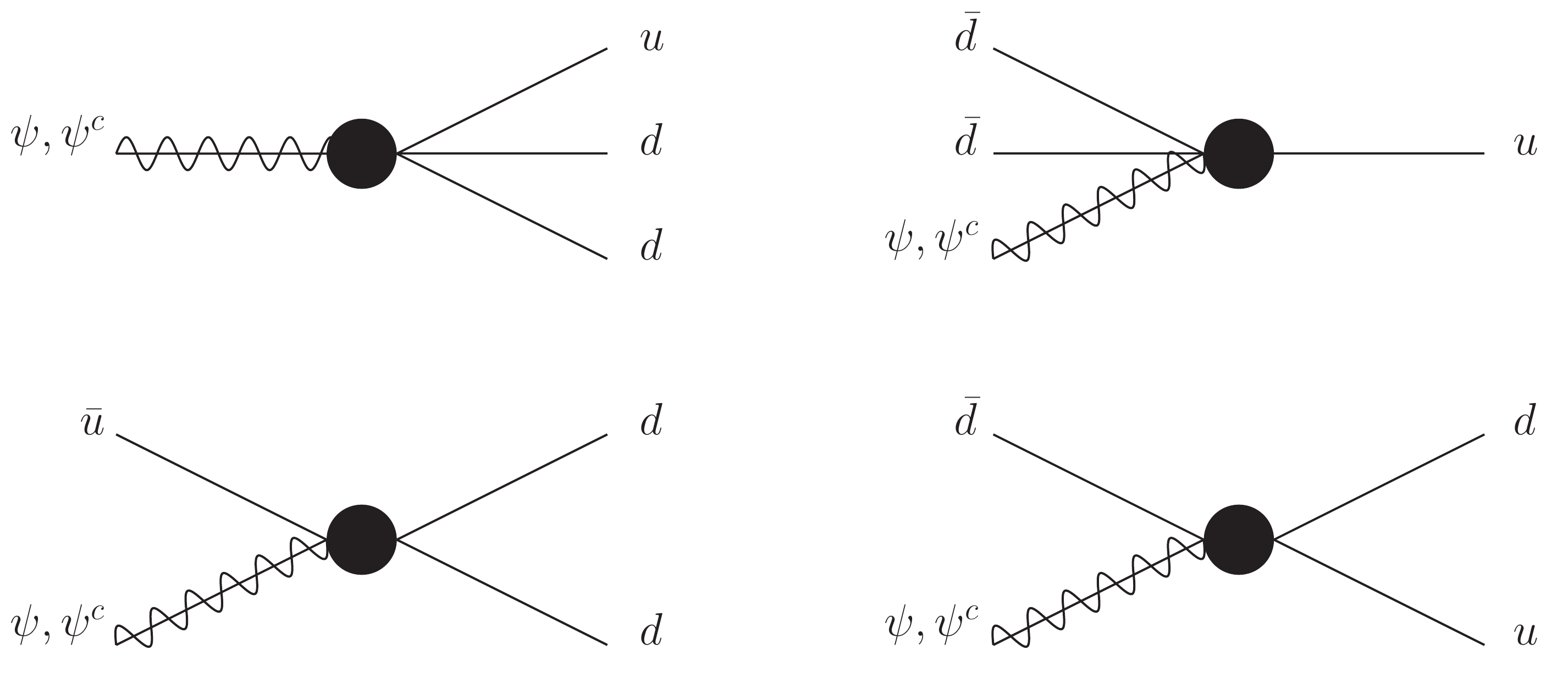}
\caption{Processes that change baryon number by one ($\Delta B=1$). The four-fermion vertices are given in Eq.\ref{eq:Leff}.}\label{fig:dB1}
\end{figure}

Now that we have a complete model, we can study baryon-number generation described in Section~\ref{sec:BAU} in more detail. In order to find the net particle number, we need the rates of processes that change that particle number. Focusing only on baryon-number-violating interactions, processes that change the baryon number by one unit in this model are shown in Fig.~\ref{fig:dB1}. 

\begin{figure}[h!]
\includegraphics[width=\linewidth]{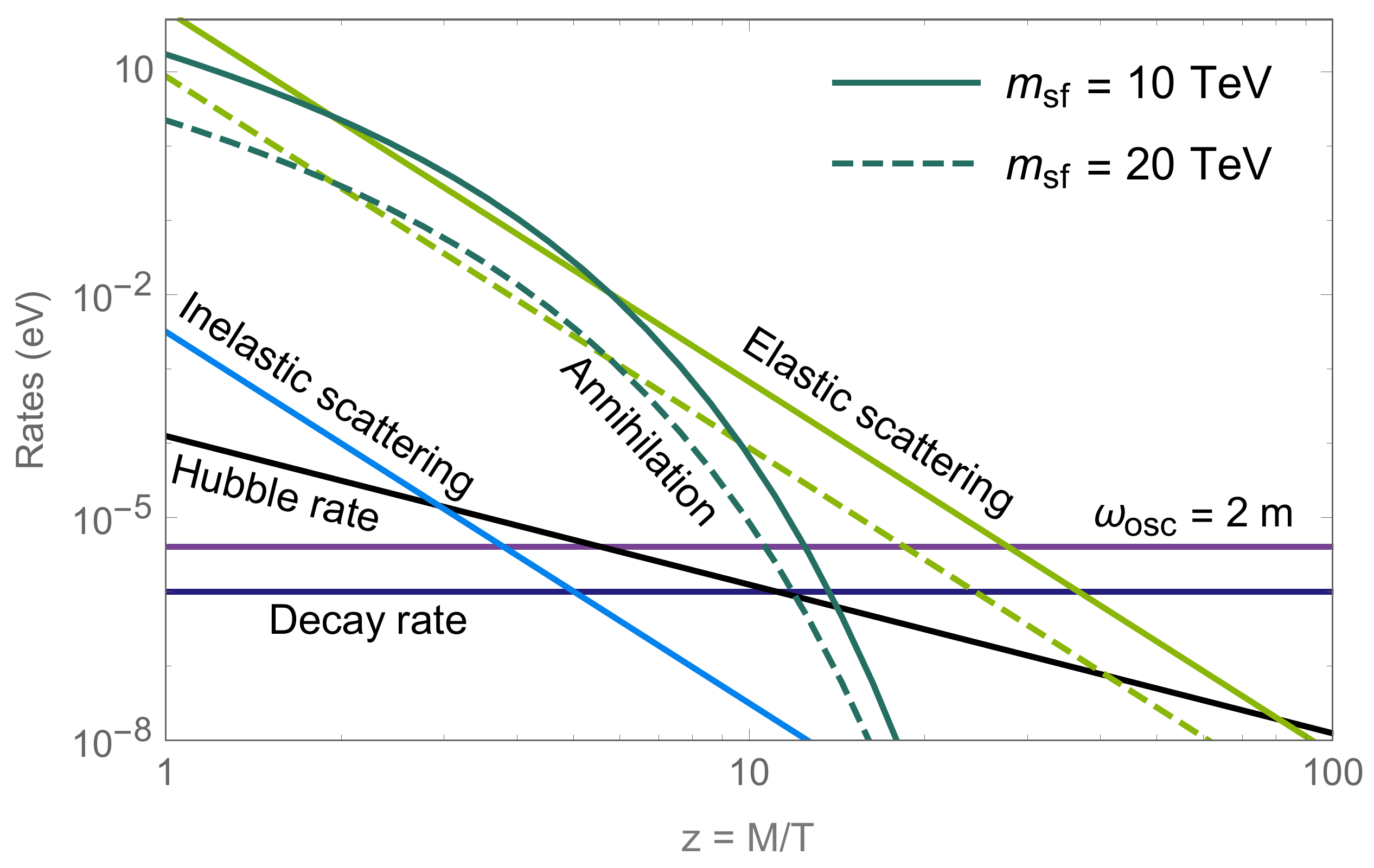}
\caption{Comparison of the decay rate, the Hubble rate, the oscillation frequency, the annihilation rate, the elastic and inelastic scattering rates for $M=300$~GeV, $m=2\times10^{-6}~$eV, $m_{\rm sq}=2~$TeV, and $\Gamma= 10^{-6}$~eV.}\label{fig:ratesvz2}
\end{figure}

As discussed in Section~\ref{sec:BAU}, baryon asymmetry is produced at low temperatures, $z=M/T\gg 10$ for squarks of mass $O(10~{\rm TeV})$. Hence we can ignore processes with $\psi/\psi^c$ in the final state, such as $dd\to\psi\bar{u}$. We can also ignore $3\to1$ processes, \emph{e.g.} $\psi dd\to \bar{u}$, since they are phase-space suppressed. Inelastic $2\to2$ scatterings such as $\psi\bar{u}\to dd$, which do not affect oscillations, happen with a rate much smaller than the decay rate of the binos for $z\gtrsim 5$ (see Fig.~\ref{fig:ratesvz2}). Hence we also ignore these scatterings. 

The only relevant processes for determining the baryon asymmetry are bino/antibino decays to a final state with baryon number $+1$ or $-1$, bino annihilations and elastic bino scatterings via vector interactions. The study in Section~\ref{sec:BAU} can be followed straightforwardly. With these approximations the relevant Boltzmann equations, for $z>z_{\rm osc}$, are 
\begin{align}
\frac{d\Sigma(z)}{dz}&\simeq-\frac{\Gamma}{zH(z)}\,\Sigma(z) -\frac{s\langle \sigma_{\rm ann} v\rangle}{2zH(z)}\biggl(\Sigma^2(z)-4Y_{\rm eq}^2\biggr), \notag \\
 \frac{d\Delta_B(z)}{dz}&\simeq\frac{\epsilon\,\Gamma}{zH(z)}\,\Sigma(z). 
\end{align}
We emphasize that the baryon asymmetry is produced only after the oscillations start, when the mass difference becomes larger than the (flavor-sensitive) elastic scattering rate, $m\gtrsim \Gamma_{\rm scat}$.
\begin{align}
z_{\rm osc}\simeq 36\left(\frac{M_D}{300~{\rm GeV}}\right)\left(\frac{10~{\rm TeV}}{m_{\rm sf}}\right)^{4/5}\left(\frac{2\times10^{-6}~{\rm eV}}{m}\right)^{1/5}.
\end{align}

\begin{figure}[h!]
\includegraphics[width=\linewidth]{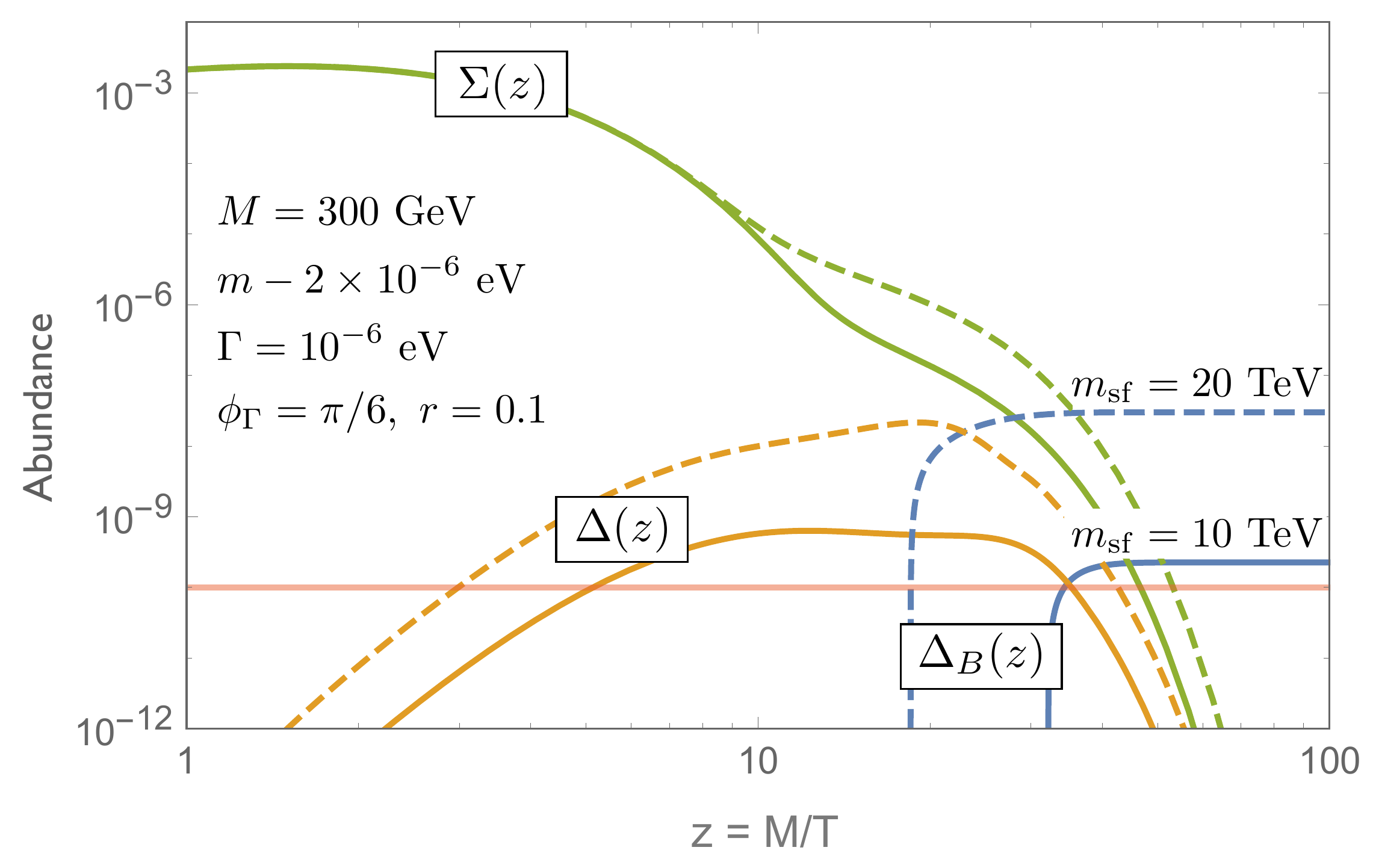}
\caption{The total $\psi$-number $\Sigma(z)$, the $\psi$-asymmetry $\Delta(z)$ and the baryon asymmetry $\Delta_B(z)$  for $m_{\rm sq}=10~$TeV (solid), 20~TeV (dashed) and $M=300~{\rm GeV},~ m=2\times 10^{-6}~{\rm eV},~\Gamma=10^{-6}~{\rm eV},
~r=0.1,~\sin\phi_\Gamma=0.5$.}\label{fig:basym2}
\end{figure}

For sfermion masses smaller than a few TeV (and Majorana masses smaller than $10^{-6}~$eV), oscillations start at $z>40$, when the bino abundance is highly suppressed. Thus it is not possible to get the right baryon asymmetry with sfermions lighter than $O(10~{\rm TeV})$. (We find that sfermions as light as 3~TeV can be accommodated in the parameter region with $\Gamma\sim 10^{-4}~$eV and $m\sim (1-10^{-2})~$eV.) Baryon asymmetries for sfermion masses 10--20~TeV are shown in Fig.~\ref{fig:basym2}.

\section{Summary and Outlook} \label{sec:summary}

In this paper we studied the oscillations of a  pseudo-Dirac fermion, $\psi$ with mass $M=300$~GeV using the density matrix description of Ref.~\cite{Tulin:2012re} that incorporates the Hubble expansion, elastic scatterings and annihilations into the time evolution of the number densities. The $\psi$-particles decay out of thermal equilibrium if their decay rate $\Gamma<H(T\sim M)\sim 10^{-4}$~eV. As benchmark values we took $\Gamma=10^{-6}$~eV and a mass difference, between the heavy and light mass eigenstates, $\Delta m=4\times 10^{-6}$~eV. We assumed that these new particles and their antiparticles were produced with a thermal number density at temperatures much higher than their mass. If there is also $CP$ violation in the system, it is enhanced for $\Gamma\sim \Delta m$. In this case, a $\psi$-asymmetry is produced at later times even if the initial densities are symmetric. Furthermore if the decays of the $\psi$-particles violate baryon number, then a baryon asymmetry can be produced. The size of the baryon asymmetry depends strongly on when the oscillations start and how they proceed in the early Universe. Here we summarize the main points of our analysis.
\begin{enumerate}
\item After being produced, particles and antiparticles cannot start oscillating right away. Even without any interactions, $\psi$-particles do not have sufficient time to oscillate before the Hubble rate drops below the oscillation frequency, $\omega_{\rm osc}=\Delta m>H(T)$. For a mass difference smaller than $10^{-6}$~eV full oscillations only start when $T\gtrsim M/10$. 

\item For electroweak scale particles that fall out of thermal equilibrium at $T\lesssim M$, interactions with light (SM) particles inhibit oscillations. For interaction cross sections larger than $O(10^{-2}~{\rm ab})$, this delay is stronger than the one due to the expansion of the Universe. If the interactions cannot differentiate between a particle and an antiparticle (flavor-blind interactions), elastic scatterings do not affect oscillations. However oscillations are delayed due to particle--antiparticle annihilations. Oscillations can be delayed until  $T\sim M/20$ for an annihilation cross-section $\sim$fb. 

\item If there are light particles that scatter off of the $\psi$-particles, and if these scatterings differentiate between a particle and an antiparticle (flavor-sensitive interactions), oscillations are delayed further. Since the elastic scattering rate is not Boltzmann suppressed (if there are light particles to scatter with), the delay due to scatterings is stronger than the delay due to annihilations. For a flavor-sensitive elastic scattering cross section $\sim$fb, oscillations can be delayed until  $T\sim M/80$. We showed the relationship between the oscillation-onset temperature and the mass difference for different interaction types and strengths in Fig.~\ref{fig:zosc}. 

\item We showed in Fig.~\ref{fig:basym} that a large baryon asymmetry can be produced if interactions that delay oscillations are not stronger than $O$(fb). For stronger interactions oscillations are usually delayed until the total $\psi$ density is too small to produce a large asymmetry even with $O(1)$ $CP$ violation. 
\end{enumerate}

As a concrete example of this scenario, we studied pseudo-Dirac bino oscillations in a $U(1)_R$-symmetric SUSY model with R-parity violation. If the lightest neutralino is a pure bino, it decays via R-parity-violating interactions. Assuming baryon-number-violating bino decays dominate we showed that when the binos decay out of thermal equilibrium, they can produce a sufficiently large baryon asymmetry to explain the baryon asymmetry of the Universe. However in order to produce enough asymmetry, sfermions need to be heavier than a few TeV lest bino oscillations are delayed too much due to strong elastic scatterings with light SM fermions. 

An important collider signature of this scenario is displaced vertices. Since these particles, with electroweak scale masses, decay out of thermal equilibrium, their decay rate $\Gamma\lesssim 10^{-4}$~eV. Consequently, if they are produced at colliders, they will travel more than a few mm before decaying. (See Ref.~\cite{Cui:2014twa}.) Furthermore, if there are lepton-number-violating decays (as well as baryon-number-violating decays), the decays can produce a same-sign lepton asymmetry \cite{Ipek:2014moa}. 

On the model building side, the oscillations can be embedded in a dark sector and be the source of the dark matter relic density together with the baryon asymmetry. As pointed out in Refs.~\cite{Cirelli:2011ac, Tulin:2012re}, one usually imposes a global $U(1)$ symmetry on the dark sector such that the dark matter particle is stable. However this global symmetry must be broken due to gravity. Then it is expected, \emph{e.g.}, the fermions in the dark sector are pseudo-Dirac particles and they undergo particle--antiparticle oscillations as described in this work. If, for example, the global symmetry is $U(1)_{B-L}$, an asymmetry that is produced by oscillations in the dark sector can be transferred into the SM baryon asymmetry.

\begin{acknowledgements}
S.I. thanks Ann Nelson and David McKeen for helpful comments on an earlier version of this manuscript. S.I. also thanks David McKeen for valuable conversations. This research has been supported in part by the Balzan foundation via New College Oxford and by the U.S. Department of Energy under Grant No. DE-SC0011637. Fermilab is operated by Fermi Research Alliance, LLC under Contract No. DE-AC02-07CH11359 with the United States Department of Energy.
\end{acknowledgements}

\appendix
\onecolumngrid
\section{Time-Dependent Decay Rates} \label{sec:timedepdec}
From Eq.~\ref{eq:Lint} we have
\begin{align}
\langle B X |-\mathcal{L}|\psi\rangle&=g_\chi, \quad \langle B X|-\mathcal{L}|\psi^c\rangle=g_\eta , \notag \\
\langle \bar{B}\bar{X}|-\mathcal{L}|\psi\rangle &=g_\eta^\ast, \quad \langle \bar{B}\bar{X}|-\mathcal{L}|\psi^c\rangle=g_\chi^\ast
\end{align}

Using Eq.~\ref{eq:timeevolv} we write the time-dependent decay rates as

\begin{align}
\Gamma(\psi(t)\to BX)&=\gamma\, |\langle B X|-\mathcal{L}|\psi(t)\rangle|^2 = \gamma \left|g_\chi g_+ -g_\eta\frac{q}{p}g_-\right|^2=\gamma\left[ |g_\chi|^2|g_+|^2+|g_\eta|^2\left|\frac{q}{p}\right|^2|g_-|^2 -2\Re\left(\frac{q}{p}g_\chi^\ast g_\eta g_+^\ast g_- \right) \right], \notag\\
\Gamma(\psi(t)\to \bar{B}\bar{X})&=\gamma\, |\langle \bar{B}\bar{X}|-\mathcal{L}|\psi(t)\rangle|^2 = \gamma \left|g_\eta^\ast g_+ -g_\chi^\ast\frac{q}{p}g_-\right|^2=\gamma\left[ |g_\eta|^2|g_+|^2+|g_\chi|^2\left|\frac{q}{p}\right|^2|g_-|^2 -2\Re\left(\frac{q}{p}g_\chi^\ast g_\eta g_+^\ast g_- \right) \right], \notag\\
\Gamma(\psi^c(t)\to BX)&=\gamma\, |\langle B X|-\mathcal{L}|\psi^c(t)\rangle|^2 = \gamma \left|g_\eta g_+ -g_\chi\frac{p}{q}g_-\right|^2=\gamma\left[ |g_\eta|^2|g_+|^2+|g_\chi|^2\left|\frac{p}{q}\right|^2|g_-|^2 -2\Re\left(\frac{p}{q}g_\eta^\ast g_\chi g_+^\ast g_- \right) \right], \notag\\
\Gamma(\psi^c(t)\to \bar{B}\bar{X})&=\gamma\, |\langle \bar{B} \bar{X}|-\mathcal{L}|\psi^c(t)\rangle|^2 = \gamma \left|g_\chi^\ast g_+ -g_\eta^\ast\frac{p}{q}g_-\right|^2=\gamma\left[ |g_\chi|^2|g_+|^2+|g_\eta |^2\left|\frac{p}{q}\right|^2|g_-|^2 -2\Re\left(\frac{p}{q}g_\eta^\ast g_\chi g_+^\ast g_- \right) \right], \notag\\
\end{align}
where $\gamma$ is determined by the details of the specific model. Then
\begin{align}
\Gamma(\psi(t)\to BX) - \Gamma(\psi(t)\to \bar{B}\bar{X})&=\gamma\, |g_\chi|^2(1-r^2)\left( |g_+|^2-\left|\frac{q}{p}\right|^2|g_-|^2 \right),\notag \\
\Gamma(\psi(t)\to BX) + \Gamma(\psi(t)\to \bar{B}\bar{X})&=\gamma\,|g_\chi|^2\left((1+r^2)\left[ |g_+|^2+\left|\frac{q}{p}\right|^2|g_-|^2 \right]-4\Re\left(\frac{q}{p}\frac{g_\eta}{g_\chi}g_+^\ast g_-\right)\right),\notag \\
\Gamma(\psi^c(t)\to BX) - \Gamma(\psi^c(t)\to \bar{B}\bar{X})&=\gamma\,|g_\chi|^2(r^2-1) \left( |g_+|^2-\left|\frac{p}{q}\right|^2|g_-|^2 \right),\notag \\
\Gamma(\psi^c(t)\to BX) + \Gamma(\psi^c(t)\to \bar{B}\bar{X})&=\gamma\,|g_\chi|^2\left((1+r^2)\left[ |g_+|^2+\left|\frac{p}{q}\right|^2|g_-|^2 \right]-4\Re\left(\frac{p}{q}\frac{g_\eta^\ast}{g_\chi^\ast}g_+^\ast g_-\right)\right).
\end{align}

We also have
\begin{align}
\frac{q}{p}\frac{g_\eta}{g_\chi}=r\qpabs e^{i\beta},\quad \frac{p}{q}\frac{g_\eta^\ast}{g_\chi^\ast}=r\pqabs e^{-i\beta}.
\end{align}
These can be combined to get Eq.~\ref{eq:epsilon}. To take the time integrals we use Eq.~\ref{eq:gpm} and get
\begin{align}
|g_{\pm}|^2&=\frac{e^{-\Gamma t}}{2}\left[\cosh\left(\frac{\Delta\Gamma}{2}t\right)\pm\cos(\Delta m\,t)\right], \notag \\
g_+^\ast g_-&=\frac{e^{-\Gamma t}}{2}\left[\sinh\left(\frac{\Delta\Gamma}{2}t\right)+ i\sin(\Delta m\,t)\right].
\end{align}

\section{Boltzmann Equations}
The Boltzmann equations from Eq.~\ref{eq:oscBoltzmann} are
\begin{align}
H z \frac{dY_\psi}{dz}&=-\Gamma\, Y_\psi +im\,(Y_{\psi\psi^c}-Y_{\psi^c\psi})-\Gamma\, r\,(e^{-i\phi_\Gamma}Y_{\psi\psi^c}+e^{i\phi_\Gamma}Y_{\psi^c\psi})  -s \langle \sigma v\rangle_+(Y_\psi Y_{\psi^c}+Y_{\psi\psi^c}Y_{\psi^c\psi}-Y_{\rm eq}^2) \notag\\
&\hspace{3.6in} -s \langle \sigma v\rangle_-(Y_\psi Y_{\psi^c}-Y_{\psi\psi^c}Y_{\psi^c\psi}-Y_{\rm eq}^2), \notag\\
H z \frac{dY_{\psi^c}}{dz}&=-\Gamma\, Y_{\psi^c} -im\,(Y_{\psi\psi^c}-Y_{\psi^c\psi})-\Gamma\, r\,(e^{-i\phi_\Gamma}Y_{\psi\psi^c}+e^{i\phi_\Gamma}Y_{\psi^c\psi}) -s \langle \sigma v\rangle_+(Y_\psi Y_{\psi^c}+Y_{\psi\psi^c}Y_{\psi^c\psi}-Y_{\rm eq}^2) \notag \\
&\hspace{3.6in} -s \langle \sigma v\rangle_-(Y_\psi Y_{\psi^c}-Y_{\psi\psi^c}Y_{\psi^c\psi}-Y_{\rm eq}^2), \notag\\
H z \frac{dY_{\psi\psi^c}}{dz}&= -\Gamma\, Y_{\psi\psi^c}+im\, (Y_\psi-Y_{\psi^c})-\Gamma\, r\, e^{i\phi_\Gamma}\,(Y_\psi+Y_{\psi^c}) -s \langle \sigma v\rangle_+ Y_{\psi\psi^c}(Y_\psi+ Y_{\psi^c})-2\Gamma_- Y_{\psi\psi^c},\notag\\
H z \frac{dY_{\psi^c\psi}}{dz}&= -\Gamma\, Y_{\psi^c\psi}-im\, (Y_\psi-Y_{\psi^c})-\Gamma\, r\, e^{-i\phi_\Gamma}\,(Y_\psi+Y_{\psi^c})-s \langle \sigma v\rangle_+ Y_{\psi^c\psi}(Y_\psi+ Y_{\psi^c})-2\Gamma_- Y_{\psi^c\psi}.  \label{eq:Boltzoscdecann}
\end{align}
\twocolumngrid

\end{document}